\documentclass[aps,twocolumn,prc,superscriptaddress,showpacs,nofootinbib,floatfix,amssymb,amsfonts,amsmath]{revtex4-1}

\usepackage{graphicx}
\usepackage{dcolumn}
\usepackage{bm}
\usepackage{xcolor}
\usepackage{amsmath}    
\usepackage{amsfonts}   
\usepackage{amssymb}
\usepackage{graphicx}   
\usepackage{multirow} 

\begin{document}

\title{Further steps towards next generation of covariant energy density functionals}

\author{B. Osei}
\affiliation{Department of Physics and Astronomy, Mississippi
State University, MS 39762}

\author{A. V. Afanasjev}
\affiliation{Department of Physics and Astronomy, Mississippi
State University, MS 39762}

\author{A. Taninah}
\affiliation{Department of Chemistry and Physics, Louisiana 
                State University, Shreveport, LA 71115.}

\author{A. Dalbah}
\affiliation{Department of Physics and Astronomy, Mississippi
State University, MS 39762}

\author{U. C. Perera}
\affiliation{Department of Physics and Astronomy, Mississippi
State University, MS 39762}

\author{V. A. Dzuba}
\affiliation{School of Physics, University of New South Wales, Sydney 2052, Australia}

\author{V. V. Flambaum}
\affiliation{School of Physics, University of New South Wales, Sydney 2052, Australia}

\date{\today}

\begin{abstract}
 
     The present study aims at  further development of covariant energy density functionals 
(CEDFs) towards more accurate description of binding energies across the nuclear chart. 
For the first time,  infinite basis corrections to binding energies in the fermionic and bosonic 
sectors of the covariant density functional theory have been taken into account in the fitting 
protocol within the covariant density functional theory. In addition,  total electron binding energies 
have been used in the conversion of atomic binding energies into nuclear ones.  Their dependence  
on neutron excess  has been investigated for the first time across the nuclear chart within atomic 
approach.  These factors have been disregarded in previous generation of covariant energy density 
functionals but their neglect leads to substantial global calculation errors for physical quantities of 
interest.  For example, these errors for binding energies are of the order of 0.8 MeV or higher for 
the three major classes of covariant energy density functionals.

 \end{abstract}

\maketitle

\section{Introduction}

    Nuclear binding energies (or masses) is one of the most fundamental properties
of atomic nuclei.  Their accurate knowledge is important for nuclear structure 
\cite{MLO.03,Eet.12,AARR.14},  neutrino physics \cite{DBBE.18} and nuclear 
astrophysics \cite{AGT.07,CSLAWLMT.21,CH.08,LBGABDG.18}. Nuclear binding
energies also provide important constraint on the properties of energy density 
functionals (EDFs) in nuclear density functional theories (nuclear DFTs) 
\cite{BHP.03,VALR.05}.

    Over the years there were substantial and continuous efforts in non-relativistic 
DFTs  to improve  global reproduction of experimental binding energies (see, for example, 
Refs.\ \cite{TGPO.00,SGHPT.02,GCP.13,BCPM,MSIS.16,UNEDF0,UNEDF1,UNEDF2}
and references quoted therein).  The global fits at the mean field level in the Skyrme 
DFT provide the $1.4 - 1.9$ MeV accuracy in the global description of  experimental 
binding energies (see Refs.\ \cite{UNEDF0,UNEDF1,UNEDF2}). The accounting 
of beyond mean field effects leads to a further improvement: the experimental binding 
energies are reproduced with an accuracy of around 0.5  and 0.8 MeV in the Skyrme 
\cite{GCP.13} and Gogny \cite{D1M} DFTs, respectively.
   
  The situation in covariant DFT (further CDFT) is more complicated due to numerical
reasons: such calculations are significantly more numerically challenging 
and more time-consuming as compared with those carried out in non-relativistic DFTs due 
to relativistic nature of CDFT (see detailed discussion in Introduction of Ref.\ \cite{TOAPT.24}). 
As a consequence only in Ref.\ \cite{AGC.16}  has an attempt to create the CEDF based on 
the global set of experimental data on binding energies  been undertaken before 2023.
It is only the advent of the anchor based optimization approach (further ABOA) in Ref.\ 
\cite{TA.23} which allowed global fits of CEDFs at a reasonable computational cost 
(see Ref.\ \cite{TOAPT.24}). In this approach, the optimization of the parameters of EDFs 
is carried out for a selected set of spherical anchor nuclei, the physical observables of which 
are modified by the correction function, which takes into account the global performance of 
EDFs. The computational cost of defining a functional within this approach is drastically (by 
at least an order of magnitude) lower as compared with alternative methods of global fits of 
EDFs (see Ref.\ \cite{TOAPT.24}). It is shown in Ref.\ \cite{TA.23}  that the use of this approach 
leads to a substantial  (by $0.5-0.8$ MeV dependent on the class of CEDF) improvement
in the global description of binding energies.

   The present paper aims at further improvement of several classes of CEDFs for the 
calculations of binding energies with high precision in the CDFT framework. The 
following improvements have been implemented in the fitting protocol of CEDFs for 
the first time.  {\it First,} infinite basis corrections for binding energies in fermionic and bosonic (mesonic) 
sectors of the CDFT have been taken into account. 
{\it Second}, total electron binding energies are taken 
into account in the extraction of nuclear binding energies from atomic binding energies
presented in Atomic Mass Evaluations (AME). 
{\it Third}, a complete expression (see Eq.\ (\ref{charge-radii}) below) for charge radii 
which also includes  spin-orbit contributions is used for anchor spherical nuclei in the 
fitting protocol.

   The paper is organized as follows. The details of the fitting protocol employed in the present 
paper are discussed in Sec.\ \ref{fitting-protocol}.  Sec.\ \ref{electron} considers the importance 
of total electron binding energies in the definition of nuclear binding energies from atomic ones 
presented in AME.  The impact of neutron excess on total electron binding 
energies is analyzed for the first time across the nuclear chart  in Sec.\ \ref{electron-binding}. 
The definition of the pseudodata on nuclear binding energies used in ABOA  and the role of 
infinite basis corrections to binding energies are discussed  in Sec.\ 
\ref{Pseudo-data}.  Sec.\ \ref{new-functionals} is dedicated to the analysis of new CEDFs defined 
in the present work. Finally, Sec.\ \ref{Concl} summarizes the  results  of our paper.

\section{Fitting protocol}
\label{fitting-protocol}

   In the present paper, numerical calculations and the fitting of CEDFs have been 
carried out in the relativistic Hartree-Bogoliubov (RHB) approach (see Refs.\ 
\cite{AARR.14,TAAR.20} for details) using anchor based optimization approach  
the technical details of which are presented  in Refs.\ \cite{TA.23,TOAPT.24}. However, 
as compared with latter two studies the  following substantial changes in the fitting protocol 
have been implemented:

\begin{itemize}
\item
   Infinite basis corrections to the binding energies due to fermionic and mesonic sectors
of CDFT have been taken into account in the fitting protocol for the first time. The 
technical details  of these corrections are discussed in detail in Sec.\ \ref{Pseudo-data}. 
Note that the largest bases used before this work
 in the optimization of  CEDFs are those in which 
fermionic and bosonic bases are truncated at full $N_F=20$ fermionic and full $N_B=20$
bosonic shells, respectively (see discussion in Sect.\ V of Ref.\ \cite{TOAPT.24} and
Ref.\ \cite{TA.23}). 

\item
Nuclear binding energies have been defined by subtracting total electron binding 
energies from atomic binding energies presented in Atomic Mass Evaluations (AMEs)  for 
the first time in CDFT (see Sec.\ \ref{electron}). 

\item
A general CDFT expression for a charge radius $r_{ch}$ given by (see Refs.\ \cite{HP.12,KT.19})
\begin{eqnarray}
r^2_{ch} = \left<r^2_p\right> + r^2_p +\frac{N}{Z} r^2_n &+& \left<r^2_p\right>_{SO} + \nonumber \\
          &+&\frac{N}{Z} \left<r_n^2\right>_{SO} 
\label{charge-radii}          
\end{eqnarray}
is used for the calculation  of charge radii of anchor spherical nuclei  in ABOA.  Here 
$\left<r^2_p\right>$  stands for point-proton mean square radius as emerging from the 
CDFT calculations, $r_p$ and $r_n$  are mean square radii of single proton and neutron, 
respectively, and $\left<r^2_p\right>_{SO}$  and $\left<r^2_n\right>_{SO}$ are proton and 
neutron spin-orbit contributions to the charge radius. The  values of $r_n^2=-0.1161$ 
fm$^2$ and $r_p=0.8409$ fm are used in the calculations (see Sec.\  IV.D of Ref.\ \cite{TOAPT.24} 
for further discussion).  Note that to our knowledge this is the first time when all terms of Eq.\ 
(\ref{charge-radii}) are used in the fitting protocol of CEDFs: early fits carried out by different 
groups are restricted to the first two terms of this equation since they dominate Eq.\ 
(\ref{charge-radii})\footnote{The impact of third term (i.e.   $\frac{N}{Z} r^2_n$) of 
Eq.\ (\ref{charge-radii})  is modest: for nuclei shown 
in Table \ref{ABOA-data} its value varies from -0.1161 fm$^2$ for $^{16}$O  to -0.1904 
fm$^{2}$ for $^{132}$Sn. Moreover, its relative contribution to  $r_{ch}^2$ decreases 
with increasing mass: this term represents only approximately 1.6\% and 0.6\% of  total 
value of  $r_{ch}^2$ in the $^{16}$O and $^{214}$Pb nuclei. However,  it still  visibly 
affects  differential charge radii in long isotopic chains because of the $N/Z$ factor.

 Proton and neutron spin-orbit contributions to the charge radii are typically smaller.
They are zero when both partners of the spin-orbit doublet are fully occupied (see 
Refs.\ \cite{HP.12,RN.21,NOSW.23}). This takes place for proton subsystem in $^{16}$O, 
$^{40,48}$Ca,  $^{90}$Zr and  for neutron subsystem in $^{16}$O and $^{40}$Ca. The 
maximum value of spin-orbit contributions to charge radii is reached when the lowest 
partner of spin-orbit doublet with $j=l+1/2$ is fully occupied but the highest one with 
$j=l-1/2$ is fully empty  \cite{HP.12,RN.21,NOSW.23}. Among the nuclei presented in 
Table \ \ref{ABOA-data}  the largest calculated values of $\left<r^2_p\right>_{SO}$ 
and $\left<r_n^2\right>_{SO}$ are 0.0679 fm$^{2}$ and -0.1207 fm$^2$ in the $^{72}$Ni
nucleus, respectively.
However, there is partial cancelation of the contributions of fourth and fifth terms of 
Eq.\ (\ref{charge-radii}) because of their opposite signs. As a result, a combined 
contribution of these two terms is always smaller than the one of the third term in 
Eq.\ (\ref{charge-radii}). Note that calculated values for proton and neutron spin-orbit 
contributions to the charge radii are for the DD-MEZ functional; however, they only 
weakly depend on the functional.  }. 
In the global calculations, only first three terms of Eq.\ 
(\ref{charge-radii}) are taken into account following suggestion of Ref.\ \cite{RN.21} 
(see also Sec.\ IV.D of Ref.\ \cite{TOAPT.24} for additional discussion of this 
approximation).

\end{itemize} 

\begin{table}[htb]
\begin{center}
\caption{ Experimental nuclear  binding energies $B$ and charge radii $r_{ch}$ 
              of 12 spherical anchor nuclei used in the fitting protocol. The data on
               nuclear binding energies are defined using Eq.\ (\ref{binding-nuclear}) from 
               atomic binding energies of Ref.\ \cite{AME2020-second}. The experimental 
               data on charge radii is taken from Refs.\ \cite{AM.13}.         
              }
\begin{tabular}{ccc}
\hline \hline
   Nucleus       &    $B$ [MeV]   &   $r_{ch}$ [fm]   \\  \hline
   $^{16}$O      &   -127.617     &    2.6991            \\  
   $^{40}$Ca     &   -342.034     &    3.4776           \\  
   $^{48}$Ca     &   -415.983     &    3.4771           \\  
   $^{72}$Ni     &   -613.414     &     N/A                 \\  
   $^{90}$Zr      &   -783.799     &    4.2694            \\  
   $^{116}$Sn    &   -988.514     &    4.6250           \\  
   $^{124}$Sn    &  -1049.790     &    4.6735          \\  
   $^{132}$Sn    &  -1102.675     &    4.7093          \\  
   $^{204}$Pb    &  -1606.938     &    5.4803          \\                    
   $^{208}$Pb    &  -1635.863     &    5.5012          \\ 
   $^{214}$Pb    &  -1662.724     &    5.5577          \\  
   $^{210}$Po    &  -1644.609     &     N/A              \\   \hline \hline
\end{tabular}
\label{ABOA-data}
\end{center}
\end{table}

   As discussed in Refs.\ \cite{TOAPT.24}, theoretical calculations of nuclear 
masses are very numerically demanding problem in CDFT.  Thus, in the present paper we 
consider three major classes of covariant energy density functionals (see  
Sec. III in Ref.\ \cite{TOAPT.24} for more details) with the goal to define the 
class of the functionals which both potentially can give the best description of masses in 
CDFT and is best suited  for the fitting at beyond mean field level. These classes of the 
functionals are (i) those based on meson exchange with nonlinear meson couplings (NLME) 
(see Ref.\ \cite{BB.77}), (ii) those based on meson exchange with density dependent 
meson-nucleon couplings (DDME) (see Ref.\ \cite{TW.99}), and finally (iii) those based 
on point coupling (PC) models containing various zero-range interactions in the Lagrangian 
(see Ref.\  \cite{PC-F1}).

  In order to avoid the uncertainties connected with the definition of the size of the 
pairing window the separable form of the finite range Gogny pairing interaction introduced in 
Ref.\ \cite{TMR.09} is used with particle number dependent strength $f$ of the pairing. Following 
the global analysis of pairing indicators carried out in Ref.\ \cite{TA.21}, the proton pairing is 
made mass dependent via 
\begin{eqnarray}
f_\pi = 1.877 (N+Z)^{-0.1072},
\label{scaling-factors-pr}
\end{eqnarray}
and neutron pairing is isospin dependent via 
\begin{eqnarray}
f_\nu = 1.208 e^{-0.674\frac{|N-Z|}{N+Z}}.
\label{scaling-factors-nu}
\end{eqnarray} 
This type of phenomenological scaling of pairing strength provides the best reproduction 
of the experimental pairing indicators (see Ref.\ \cite{TA.21}). 

   The fitting protocols employed are similar to those used in the optimization 
of the DD-MEY, NL5(Y) and PC-Y1 CEDFs (see Table IV of Supplemental Material to Ref.\ 
\cite{TA.23}). They are based on experimental data on binding energies and charge radii.
The experimental data (nuclear binding energies, charge radii) of 12 anchor spherical 
nuclei used in ABOA are summarized in Table \ref{ABOA-data}. The adopted errors are
$\Delta E = 1.0$ MeV for binding energies in the NLME, DDME and PC models,
$\Delta r_{ch} = 0.002\,\, r_{ch}$ for charge radii in the NLME and DDME models,
and $\Delta r_{ch} = 0.02$ fm for the PC model.  All  882 even-even nuclei located
between proton and neutron drip lines for which experimental binding energies exist in 
the AME2020 mass evaluation are used in the definition of the correction function (see Eq. 
(2) in Ref.\ \cite{TA.23}) of ABOA.  No  constraints on nuclear matter properties are used 
in the present paper.  Note that the functionals fitted in the present paper contain letter 
"Z" at the end of their labels.

\section{The issue of nuclear binding energies in energy density
              functional fits}
\label{electron}

   Unfortunately, the fact that  binding energies $B^{AME}(Z,N)$ presented in the 
Atomic Mass Evaluations (further AME) are {\it atomic}\footnote{In the AME compilations
the binding energies (in early editions) or binding energies per nucleon (in later
editons) are extracted from {\it atomic} masses. Because of this fact, it is logical 
to call them as {\it atomic} binding energies and we follow this notation in the 
present paper.} ones has been overlooked in many fitting protocols of EDFs. 
The connection between nuclear binding energies $B(Z,N)$\footnote{$B(Z,N)$ 
is  the (negative) binding energy of the nucleus with $Z$ protons and $N$ 
neutrons. Note that some publications show binding energies as positive
quantities but binding energies are typically used as negative quantities
in theoretical nuclear physics (see, for example, Refs.\ 
\cite{BRRM.00,HBGSS.02,AARR.14,BSaad.16}).
In the present paper  we follow the convention that nuclear and atomic binding 
energies are negative.
}
and atomic ones is given 
via (see appendix  to Ref.\ \cite{LPT.03}) 
\begin{eqnarray}
B(Z,N) = B^{AME}(Z,N) + ZB_{el}(Z=1) - B_{el}(Z) 
\label{binding-nuclear}
\end{eqnarray}
where $B_{el}(Z)$ is the total electron binding energy in the atom with 
$Z$ protons.
Note that the binding energy of electron in hydrogen atom is small (13.6 $e$V) 
on nuclear scale. Thus, the second term in this expression is ignored in some 
publications.

   The experimental binding energy $^{208}$Pb quoted in a publication provides 
a simplest fingerprint on whether atomic or nuclear binding energies are used in the
fit of the functional. This is because this nucleus is used in the most of the  EDF fits
and, in addition, its atom has appreciable total electron binding energy
$B_{el}= - 0.5682$  MeV. Its atomic binding energy per nucleon is $-7867.453$ keV in 
AME2020  (see page 030003-61 in Ref.\ \cite{AME2020-second}).  Then atomic
binding energy of the $^{208}$Pb atom is  
$B^{AME}(82,126)=-7.867453\,\,{\rm MeV} \times 208\,\,{\rm nucleons}= 
 -1636.4302\,\,{\rm MeV}$ (see Table \ref{B-energy-issue}). However, 
according to Eq.\ (\ref{binding-nuclear}) nuclear binding energy of the $^{208}$Pb 
nucleus is  $ B(82,126) = -1636.4302\,\,{\rm MeV} - 82\times 0.0000136\,\,{\rm MeV}
 - (- 0.5682\,\,{\rm MeV)} = -1635.863$ MeV (see Table \ref{B-energy-issue} 
and similar  discussion in Sec.\ II.C.2 of Ref.\ \cite{UNEDF0}).

\begin{table}[htb]
\begin{center}
\caption{Nuclear and atomic binding energies [(total) binding energy $B$ and binding 
              energy per nucleon $B/A$] of $^{208}$Pb.
\label{B-energy-issue}
}
\begin{tabular}{ccc}
\hline \hline
                            &    $B$                      &    $B/A$       \\  
                            &  [MeV]                     &   [MeV/nucleon]         \\ \hline                       
  Atomic (AME)    &   -1636.4302           &    -7.867        \\  
    Nuclear           &    -1635.863             &    -7.865        \\  \hline \hline
\end{tabular}
\end{center}
\end{table}

  Based on the values of experimental binding energy of $^{208}$Pb and the details of the 
formalism  provided in respective publications one can conclude that  the functionals listed 
in the left column of Table \ \ref{B-energy-issue-compar} are fitted to atomic binding energies 
(i.e. total electron binding energies are ignored) while the ones shown in the right column to 
nuclear ones.

\begin{table*}[htb]
\begin{center}
\caption{The columns labelled as "Atomic" and "Nuclear" indicate
whether atomic or nuclear binding energies were used in the fit of indicated 
EDFs. Respective references to the functionals are provided. 
\label{B-energy-issue-compar}
}
\begin{tabular}{cc}
\hline \hline
  Atomic                                                                                     &    Nuclear      \\  \hline \\
\multicolumn{2}{c}{Covariant EDFs}     \\ \\
   NL-Z \cite{NLZ}, NL3 \cite{NL3}, NL3* \cite{NL3*},              &                    \\                    
 DD-ME1 \cite{DD-ME1}, DD-ME$\delta$ \cite{DD-MEdelta},      &                                                             \\  
 PK1, PK1R, PKDD \cite{PK1-PK1R-PKDD},                              &                      \\ 
 PC-PK1 \cite{PC-PK1},  DD-LZ1 \cite{DD-LZ1},                 &                      \\ 
 DD-ME2 \cite{DD-ME2},  no-name PC \cite{NHM.92},         &                      \\ 
   G3 \cite{G3}, DD-MEX \cite{TAAR.20},                             &                       \\
   NL5(Y), DD-MEY, PC-Y  \cite{TA.23},                                       &                      \\ 
    BSP, IUFSU*  \cite{BSP-IUFSU*},                                           &                      \\ 
    SVI-1, SVI-2  \cite{SVI1.SVI-2}, PC-3LR \cite{PC-3LR}   &                      \\ \\
%
\multicolumn{2}{c}{Skyrme EDFs}            \\  \\
%
          SII, SIII, SIV, SV, SVI \cite{SII-SVI}                                    &     UNEDF0 \cite{UNEDF0}, UNEDF1\cite{UNEDF1}              \\ 
          SV-family [10 EDFs] \cite{SV-skyrme-family}                             &     UNEDF2 \cite{UNEDF2}                                                       \\ 
          SLy4, SLy5, SLy6, SLy7 \cite{SLy4-SLy7}                                 &     LO, NLO, N2LO -    \\ 
           SkM* \cite{SkM-SkM*}                                                               &     - family [8 EDFs] \cite{LO-NLO-N2LO.18}      \\         \\
%
\multicolumn{2}{c}{Gogny EDFs}                                                                          \\   \\
%
     D1 \cite{D1},     D1S \cite{D1S-a}                                           &                      \\  \\
%
\multicolumn{2}{c}{Microscopic +macroscopic model}                                                                          \\     \\
%
    &  model of M{\"o}ller-Nix \cite{MNMS.95,MSIS.16}                                                \\  \\
%
\multicolumn{2}{c}{Weizs{\"a}cker-Skyrme nuclear mass tables}                                                                          \\   \\
 WS \cite{WS-mass-formula},  WS* \cite{WS*-mass-formula}, WS3 \cite{WS3-mass-formula} , WS4 \cite{WS4-mass-formula}  &       \\    \hline      \hline
\end{tabular}
\end{center}
\end{table*}

\section{Total electron binding energies of atoms built on nuclei across the nuclear chart.}
\label{electron-binding}
 
  For a long time, the calculations of total electron binding energies were available only for 
the $Z=2-106$ atoms \cite{HACCM.76,Santos}. It is only in 2024 that such calculations were 
extended by us  to the atoms of superheavy nuclei with proton number $Z$ up to 120 in Ref.\ 
\cite{DFA.24}. Our calculations of total electron binding energy of atoms begin with 
the relativistic Hartree-Fock (RHF) method and follow the approach presented in  
Refs.\ \cite{Breit,QED,DBHF.17,LDF.18,LDF.19,LDF.20,DFA.24}. 
Note that they are more complete and consistent than those presented in Refs.\
 \cite{HACCM.76,Santos}   since they include correlation 
corrections and electron core relaxation effect in the QED self-energy term missing in Ref.\ 
\cite{Santos}. In addition, electron exchange interaction was approximated by a local potential
in Ref.~\cite{HACCM.76}.  Moreover, Ref.\ \cite{DFA.24}  provides the assessment of theoretical 
uncertainties in the calculations of total electron binding energies.

\begin{figure}[htb]
 \begin{center}
   \centering
   \includegraphics*[width=8.5cm]{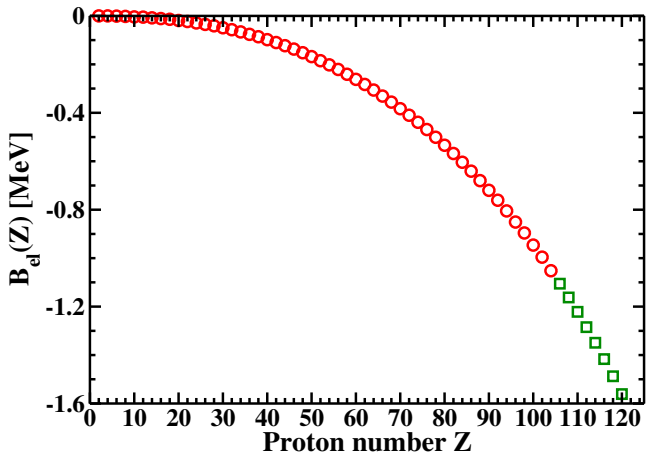}
   \caption{The evolution of total electron binding energies $B_{el}$ of the 
atoms as a function of proton number $Z$ as defined from the results presented 
in Ref.\ \cite{DFA.24}. The atoms for which earlier estimates of $B_{el}$ are available
in Refs.\ \cite{HACCM.76,Santos} are shown by open red circles. Open green 
squares show the atoms of superheavy elements for which total electron binding energies 
were calculated only in Ref.\ \cite{DFA.24}.   
\label{corr-energy-new} 
}
\end{center}
\end{figure}

  Fig.\ \ref{corr-energy-new}  shows the evolution of total electron binding energies $B_{el}$ 
of the atoms as a function of proton number $Z$.
They are defined from the results presented  in  the last column of Table III 
of  Ref.\ \cite{DFA.24} which cover the $Z=2-120$ atoms. One can see that the magnitude of 
$B_{el}$ drastically increases with increasing $Z$ and becomes especially large in superheavy nuclei 
where it exceeds 1 MeV and almost reaches 1.6 MeV for $Z=120$. This clearly indicates the 
need for accounting of total electron binding energies in accurate global mass calculations. 
 Note that Eq. (19) and Table IV of Ref.\ \cite{DFA.24} provides fitted expression 
for total electron binding energies presented in the last column of Table III of this reference
which can also be used in the fitting protocols of EDFs.

   In the context of global studies of nuclear binding energies one should also 
 understand how the total electron binding energies change with increasing nuclear charge 
radii caused by the raise of neutron number $N$ in isotopic chains. This topic  was not 
investigated in the past. Thus,  atomic calculations based on the formalism of 
Ref.\ \cite{DFA.24} are carried out in the present paper to cover this gap in our knowledge.

   Total electron binding energies are different for different isotopes due to the difference in 
nuclear charge radius  (field shift) and nuclear mass (mass shift). It is well  known that for heavy 
atoms field  shift strongly dominates (see, for example,  Ref.\ \cite{Sobelman}).
It can be approximated by 
\begin{eqnarray}
\Delta B_{el} = F \delta \langle r^2 \rangle,
\label{e:FIS}
\end{eqnarray}
where $F$ is the field shift constant and $\delta \langle r^2 \rangle$ is the change of  the square  of  
nuclear root mean square (RMS) charge  radius.  This equation allows to find isotopic shift in total 
electron binding energies  for any pair of isotopes provided that $F$ and nuclear RMS radii are know. 
Field shift constants $F$ were calculated  for different isotopic chains by varying nuclear radius in the 
atomic calculations and using  $F= \Delta B_{el}/ \delta \langle r^2 \rangle$ (see 
Table 1 in Supplemental Material).  These constants gradually increase with $Z$ and become 
especially large in superheavy nuclei.

However, only small changes  in total electron binding energy are expected with increasing 
neutron number:  the assumption of  $\delta \langle r^2 \rangle$ = 1~fm$^2$ leads to $\Delta B_{el}/B_{el} \sim 0.01\%$.
 Let us consider the Og atoms as an example. For such atoms $F=0.118$ keV/fm$^2$
 and  proton RMS radii of the $^{290}$Og and $^{314}$Og isotopes are 6.204~fm 
 and  6.570~fm, respectively (see Table \ref{table-neutron-rich}).  Then the difference in total electron binding 
 energies between these isotopes is   0.552~keV  which  is only 0.04\% of total electron binding energy.  
 Table \ref{table-neutron-rich} summarizes the results for the Pb, Fm and Og isotopic chains.  The $\Delta B_{el}$ 
 values of atoms below Pb, defined for isotopes located at neutron and proton drip lines, decrease fast with 
 decreasing $Z$. However, similarly defined $\Delta B_{el}$ values of high $Z$ superheavy nuclei can exceed 1 keV. 
 This is because such nuclei have high $F$ values (see Table 1 in Supplemental Material) and 
 $\delta \langle r^2 \rangle$ between  isotopes located at the neutron and proton drip lines can substantially 
 exceed 1~fm$^2$ since neutron drip line  is located near or  beyond $N=258$ (see Ref.\ \cite{TAA.20}).

    Thus, one can conclude that the variation of $B_{el}$ along the isotopic chain can be 
 ignored in global nuclear  mass calculations since $\Delta B_{el}$ is safely below 1 keV for 
 all isotopic chains in which experimental data on binding energies exists. As a consequence, the
 $B_{el}$ values provided in the last column of Table III of Ref.\ \cite{DFA.24} are used in the present 
 paper for total electron binding energies. 

\begin{table*}[htb]
\centering
\caption{The change of total electron binding energies $\Delta B_{el}= B_{el}(A_1) - B_{el}(A_2)$ 
on transition from the isotope with mass $A_1$ to the isotope with mass $A_2$ in indicated 
isotopic chains. The isotope with mass $A_1$ is located at proton drip line. The isotope with  
mass $A_2$ in the isotopic chain of  Pb is located at two-neutron drip line and those in the Fm 
and Og chains at $N=196$.  Field shift constants $F$ are shown in the second column.
Third column provides information on proton root-mean-square 
radii of the isotopes  which  are taken from supplemental materials of Refs.\ 
\cite{AARR.14,AANR.15}.
}
\begin{tabular}{ccccc}
\hline \hline
$^{A_2-A_1}$Nucleus   &  $F$  & $r_{rms}^p(A_2)$, $r_{rms}^p(A_1)$  & $B_{el}(A_1)$  & $\Delta B_{el}$    \\ 
                        & (keV/fm$^2$)   &   [fm]                                                    &  [keV]               &    [keV]                \\ \hline 
$^{266-176}$Pb           &  0.0041 & 5.844, 5.292                                              &        -567.98355         &  0.025                       \\
$^{314-290}$Fm          &   0.022 & 6.206, 5.882                                             &        -946.41249         &   0.086                    \\
$^{314-290}$Og           &  0.118 & 6.570, 6.204                                              &       -1487.20663       & 0.552                  \\
\hline \hline
\end{tabular} 
\label{table-neutron-rich}
\end{table*} 

\section{Pseudodata on nuclear binding energies}
\label{Pseudo-data}
 
    At present, global optimizations of the CEDFs in the fermionic basis 
which accurately approximates the behavior  of the one with infinite size is 
impossible (see Ref.\ \cite{TOAPT.24}). Thus, similar to other
recent studies (see Ref.\ \cite{TOAPT.24} for more details) the optimizations 
of the functionals  in the present paper are carried out in the RHB calculations 
with fermionic basis truncated at full $N_F=20$ fermionic shells. However, the 
fitting of the parameters is performed to the pseudodata  on nuclear binding 
energies  which takes into account infinite basis corrections in the fermionic 
sector of the CDFT  and removes the contribution of the total electron binding 
energy $B_{el}$ from atomic binding energies $B^{AME}(Z,N)$ of the Atomic 
Mass Evaluations (AME).

   The pseudodata on nuclear binding energies is defined as 
\begin{eqnarray} 
B^{pseudo} (Z,N) = B^{AME}(Z,N) - \Delta B^{cor} (Z,N), \nonumber \\
\end{eqnarray}
where atomic binding energies $B^{AME}(Z,N)$ are taken from Ref.\ 
\cite{AME2020-second}, and
\begin{eqnarray}
\Delta B^{cor} (Z,N) = B_{el}(Z) + \Delta B_{\infty}^{F}(Z,N)
\label{correct-func}
\end{eqnarray}
is the correction function.  It takes into account  total electron binding energies $B_{el}(Z)$ 
defined in Ref.\ \cite{DFA.24}  and infinite basis corrections to nuclear binding energies 
$\Delta B_{\infty}^{F}(Z,N)$ for infinite size of the fermionic ($N_F=\infty$)  basis. The latter is 
formally defined as 
\begin{eqnarray}
\Delta B_{\infty}^{F}(Z,N) = B(N_F=\infty)(Z,N) - B(N_F=20)(Z,N) \nonumber \\
\end{eqnarray}
but since numerical calculations in the infinite fermionic basis are impossible 
the procedure described in Sec. VI of Ref.\ \cite{TOAPT.24} is used for the 
definition of $\Delta B_{\infty}^{F}(Z,N)$  at each round of ABOA 
employed in the fitting protocol.  Note that $\Delta B^{cor} (Z,N)$ is  the sum of the 
quantities  presented in Figs.\ \ref{corr-energy-new} and  \ref{corr-energy-ferm-global}.

\begin{figure*}[htb]
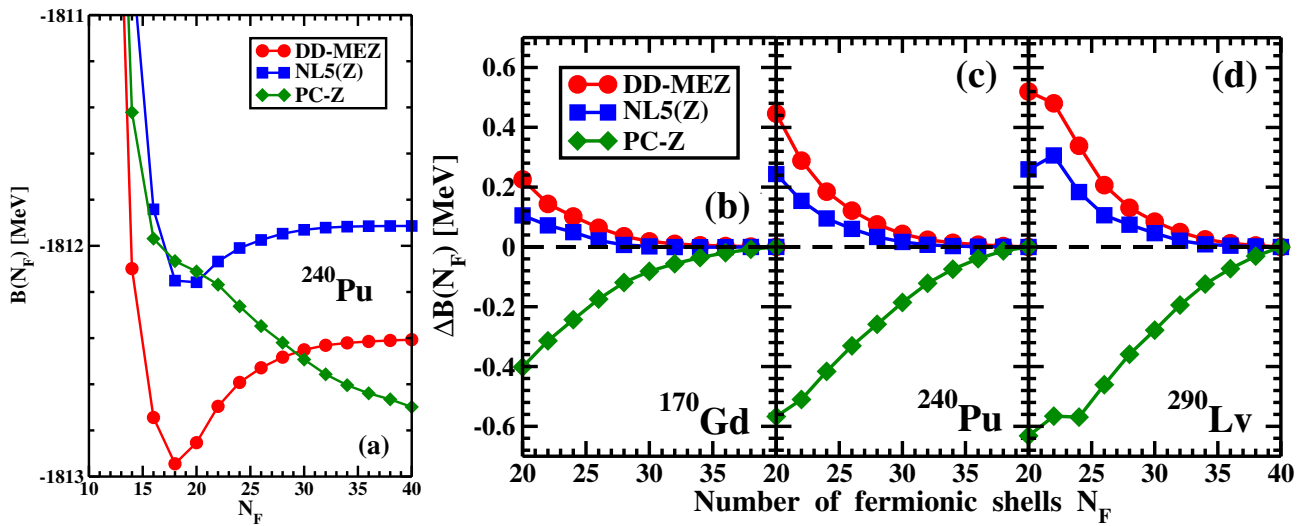

	\centering
        \includegraphics*[width=5.5cm]{fig-2-a.eps}        
        \includegraphics*[width=11.5cm]{fig-2-b.eps}        
\caption{(panel (a)). The convergence of calculated binding energies $B$ of the lowest in 
energy solution in $^{240}$Pu as a function of the number of fermionic shells $N_F$ for indicated 
functionals. (panels (b),(c) and (d)). The evolution of  $\Delta B(N_F) = B(N_F=40)-B(N_F)$  as a function 
of $N_F$ for the $^{170}$Gd, $^{240}$Pu and $^{290}$Lv nuclei. Full convergence is reached at 
$N'_F$ when $\Delta B(N'_F) = \Delta B(N'_F+2) =0$.
\label{290-LV-and-others} 
}
\end{figure*}

       It is important to benchmark the results of above mentioned procedure with 
respect of numerically exact results. However, they were never calculated in the CDFT framework 
for  medium and heavy mass nuclei because of low $N_F$ values used in the calculations 
of deformed nuclei:  the highest value of $N_F$ so far employed was $N_F=30$ in Ref.\ 
\cite{TOAPT.24} but even it  does not guarantee full convergence. The rest of existing CDFT 
calculations uses significantly smaller $N_F$ (see Sec. V of  Ref.\ \cite{TOAPT.24} for a 
survey).

\begin{figure}[htb]
  \begin{center}
    \centering
\includegraphics*[width=8.5cm]{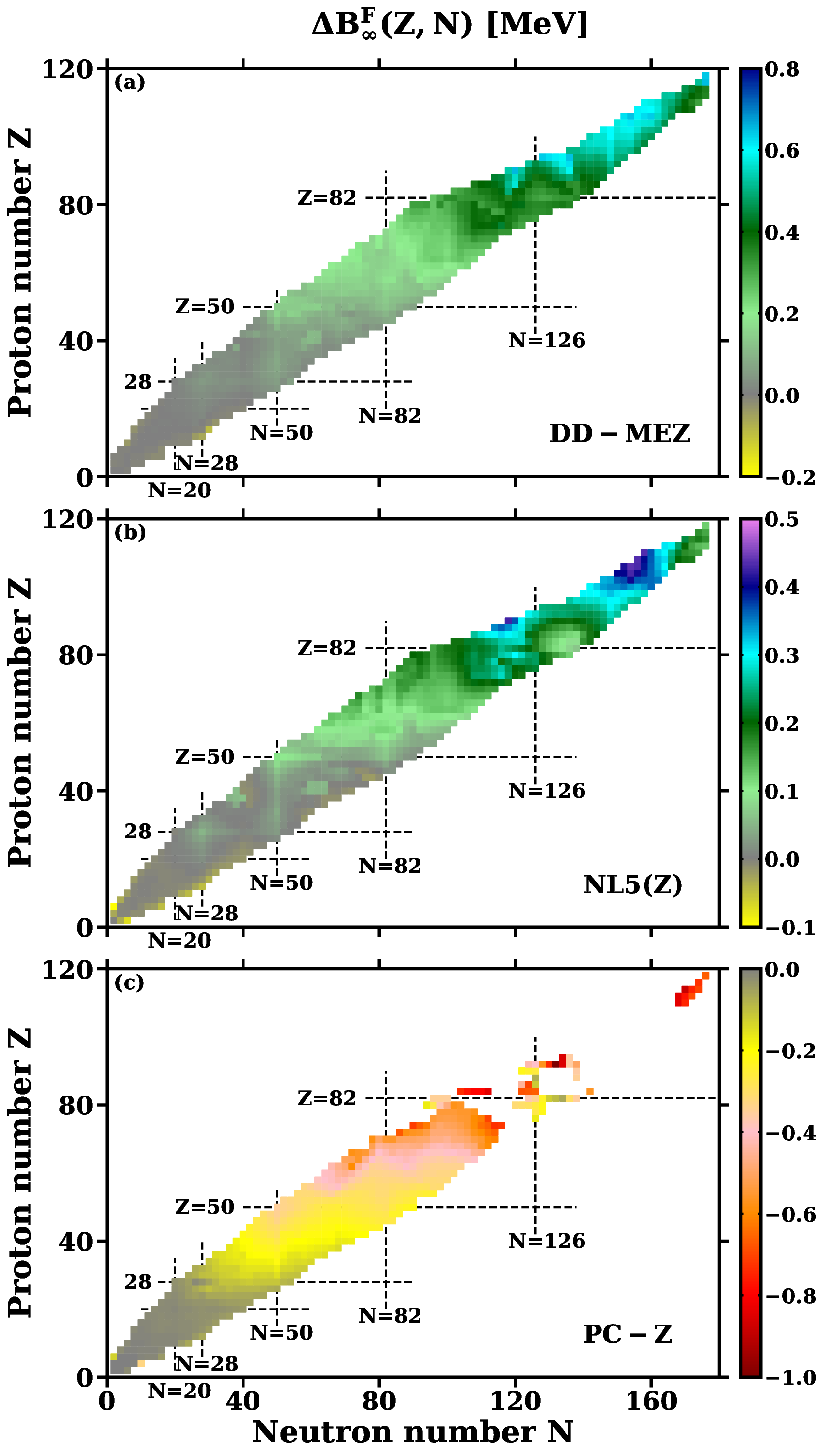}
\caption{Global maps of infinite basis corrections $\Delta B_{\infty}^{F}(Z,N)$ to nuclear  
binding energies  for the fermionic  sector of CDFT presented for indicated functionals.  In 
the case of the PC-Z functional, only the nuclei for which infinite basis corrections to nuclear 
binding energies can be defined in the fermionic basis are shown here (see also Sec. VI of 
Ref.\ \cite{TOAPT.24} for the discussion of  the PC functionals). 
\label{corr-energy-ferm-global}
}
\end{center}
\end{figure}

To address this issue the calculations with significantly modified RHB code for axially 
deformed nuclei   were carried  out for selected set of nuclei: with this code the calculations 
in the fermionic bases with $N_F$ up to 40 (with pairing) and 60 (without pairing) become 
possible for the first time.  Fig.\ \ref{290-LV-and-others}  shows that in the case of DD-MEZ 
and NL5(Z) functionals full convergence is reached at  $N_F=40$ even for heaviest nuclei 
of interest when the  pairing is taken into account.

  The convergence  of binding energies with 
$N_F$ is significantly  slower for the PC-Z  functional:  no convergence is reached even at 
$N_F=40$ for the $^{240}$Pu  and $^{290}$Lv nuclei.  It turns out that the full convergence of 
binding energy of these two nuclei is reached only at $N_F=56$ in the calculations without pairing. 
Similar problem  exists also for the DD-PC1 and PC-PK1 functionals.  It is the consequence of the 
fact that the expansion in harmonic oscillator basis states convergences  very  slowly for 
zero-range forces (see Ref.\ \cite{CAKKMV.12}).

The RHB calculations with $N_F=40$ are  by  two (one) orders of magnitudes numerically 
more  time consuming than those with $N_F=20$ ($N_F=30$) and require drastic increase of 
memory.  As a result, such calculations can be performed for only selected set of nuclei.  This is a 
reason for the use of the procedure described in Sec. VI of Ref.\ \cite{TOAPT.24}  for the  definition 
of $\Delta B_{\infty}^{F}(Z,N)$ in global calculation since it is substantially less numerically and 
memory demanding.  At the same time, this procedure reproduces fully convergent $N_F=40$ 
results obtained with the DD-MEZ and NL5(Z) CEDFs for a testing set of nuclei  scattered  more 
or less equally across experimentally known nuclear  landscape with  global accuracy better than 15 and 10 
keV, respectively.

      The present analysis indicates that infinite basis corrections to binding
energies in the fermionic basis substantially depends on the variation of the functional
(especially in medium and heavy nuclei, see Fig. 2 in Supplemental Material and its 
discussion). Thus, for the first time  they have been calculated iteratively
within the fitting protocols of the Z type of the functionals employing the procedure discussed in Sec.\ III of 
Supplemental Material.  These corrections are shown in Fig.\ \ref{corr-energy-ferm-global} 
and they  differ substantially from those presented in Fig. 10 of Ref.\ \cite{TOAPT.24} 
for the DD-MEX, NL5(E) and PC-PK1 functionals (see Fig. 3  in Supplemental Material and 
its discussion). 

\begin{figure}[htb]
\begin{center}
\centering
\includegraphics*[width=8.5cm]{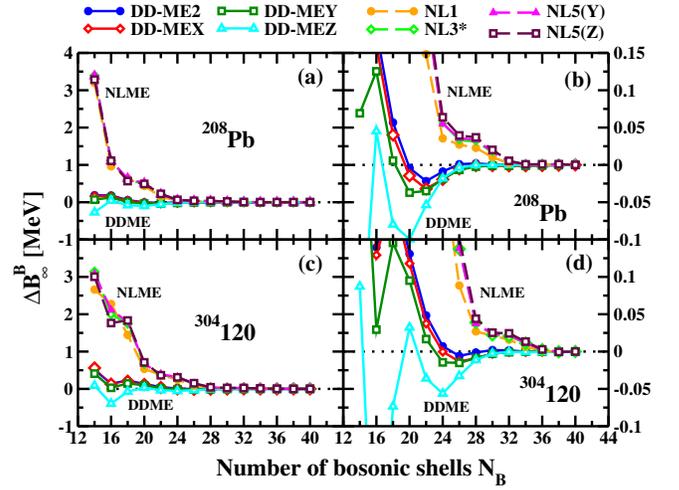}
\caption{Infinite basis corrections to nuclear binding energies 
$\Delta B^{B}_{\infty}(Z,N) = B(N_B=\infty)(Z,N) - B(N_B)(Z,N)$ for the bosonic 
sector of indicated DDME and NLME functionals as a function of $N_B$. The
results are shown for spherical $^{208}$Pb and $^{304}$120 nuclei. The right 
panels show the results presented in the left ones but in significantly reduced 
energy window. Note that the $N_F=20$ HO shells is used for fermionic basis 
in these calculations.
\label{corr-energy-bos-spher}
}
\end{center}
\end{figure}

\begin{figure}[htb]
  \begin{center}
    \centering
\includegraphics*[width=8.5cm]{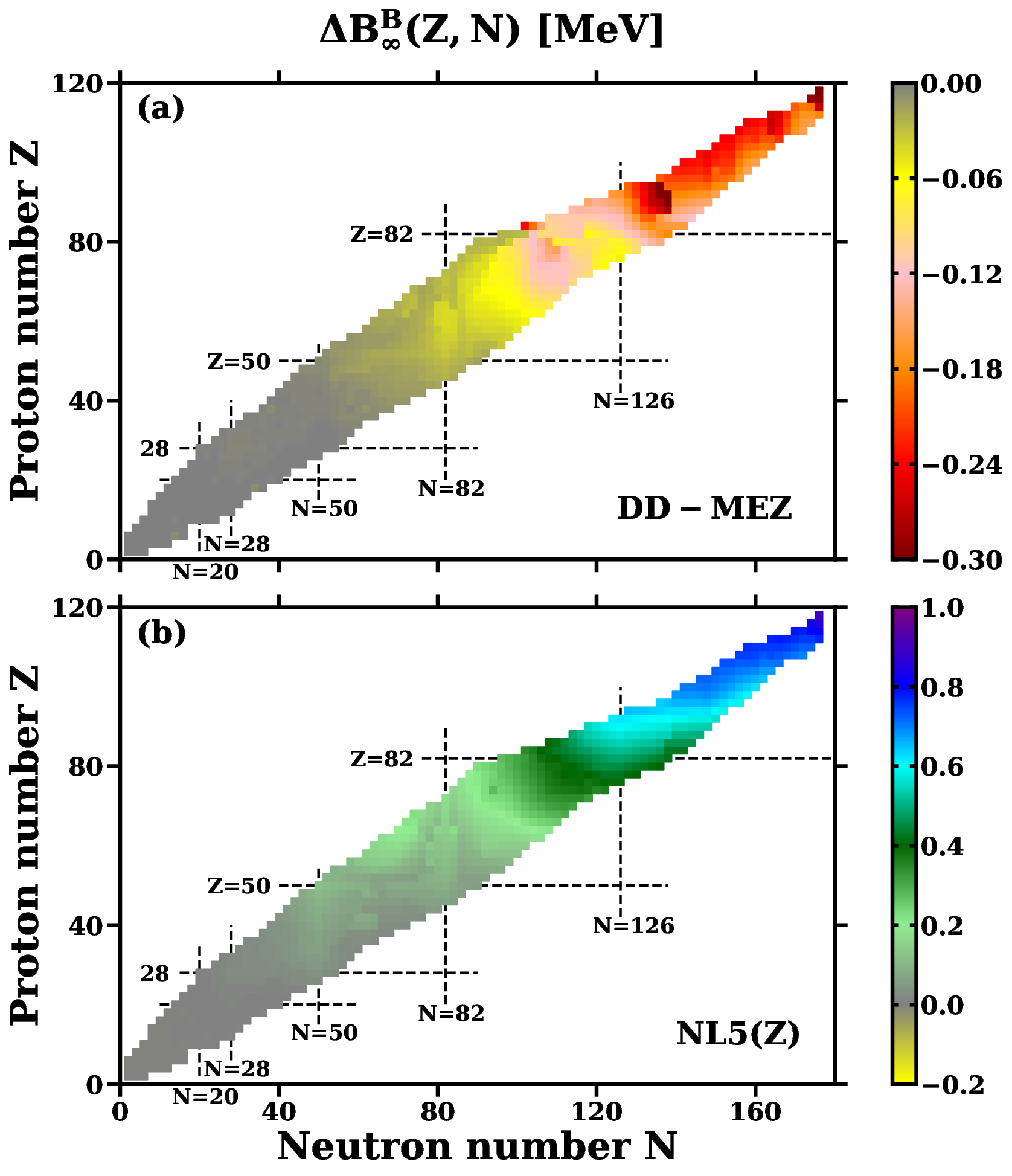}
\caption{Global maps of infinite basis corrections $\Delta B_{\infty}^{B}(Z,N)$ to nuclear 
binding energies  for the bosonic  sector of CDFT presented for the DD-MEZ and NL5(Z) 
functionals. Note that the approximation $B_{\infty}^{B}(Z,N) = B(N_B=40)(Z,N)$ is
used (see text for detail).
\label{corr-energy-bos-global}
}
\end{center}
\end{figure}

   Infinite basis corrections $\Delta B_{\infty}^{B}(Z,N)$ to nuclear binding 
energies in the bosonic sector are defined as 
\begin{eqnarray}
\Delta B_{\infty}^{B}(Z,N) = B(N_B=\infty)(Z,N) - B(N_B=20)(Z,N). \nonumber \\
\end{eqnarray}
Note that these corrections are defined for  bosonic basis with full $N_B=20$ HO 
shells which is standard value used in the CDFT calculations for over few decades.
The numerical calculations in the bosonic sector of CDFT are substantially less numerically
time and memory consuming as compared with those in the fermionic one. Thus, in the case 
of spherical nuclei we were able to carry out the RHB calculations with bosonic basis of $N_B=120$. 
The comparison with the results obtained in smaller bosonic basis indicates that full 
convergence in bosonic sector is reached at $N_B=40$ or lower dependent on the class
of the functional.  This is illustrated in Fig.\ \ref{corr-energy-bos-spher} for a number of
CEDFs representing NLME and DDME classes of the functionals.  The convergence
of the binding energies as a function of $N_B$ depends on the class of functional.
The fastest convergence is obtained for the DDME class of the functionals. In contrast,
larger bosonic basis is required for a full convergence of the NLME functionals.

  To eliminate the need for calculation of  infinite basis corrections to nuclear binding energies 
in the bosonic sector  the bosonic basis with  $N_B=40$ full HO shells is used in the present 
RHB calculations. The computational cost of such an increase (as compared with $N_B=20$) 
is small and such basis provides a numerical accuracy (as compared  with infinite bosonic basis) 
in the calculation of binding energies which is better than 10 keV for all nuclei of interest. Note 
that in light nuclei the numerical accuracy is around one keV and it gradually decreases with 
increasing mass number $A$ and reaches $\approx 10$ keV only in superheavy deformed 
nuclei.

  The importance  of such an approach to global calculations of binding energies is illustrated 
in  Fig.\  \ref{corr-energy-bos-global}. Infinite basis corrections in bosonic sector are very 
small for the $Z<50$ nuclei in the DD-MEZ functional and for the $Z<28$ nuclei in the NL5(Z) 
one  but  they gradually increase (in absolute value) with increasing proton number $Z$.  At the 
upper end of the region of superheavy nuclei they are in the vicinity of 
$\Delta B_{\infty}^{B}(Z,N) \approx -0.3$ and  $\Delta B_{\infty}^{B}(Z,N) \approx 1.0$ MeV for 
the DD-MEZ and NL5(Z) functionals, respectively.  Note that in some parts of nuclear chart the 
$\Delta B_{\infty}^{B}(Z,N)$ values show fluctuating behavior as a function of $Z$ and $N$.  
Thus, the use of $N_B=20$ in the  calculations of binding energies may lead to chaotic component 
emerging from the truncation of the bosonic basis.

\section{Discussion of new generation of the functionals}
\label{new-functionals} 

\begin{figure*}[htb]
	\centering
    \includegraphics*[width=18cm]{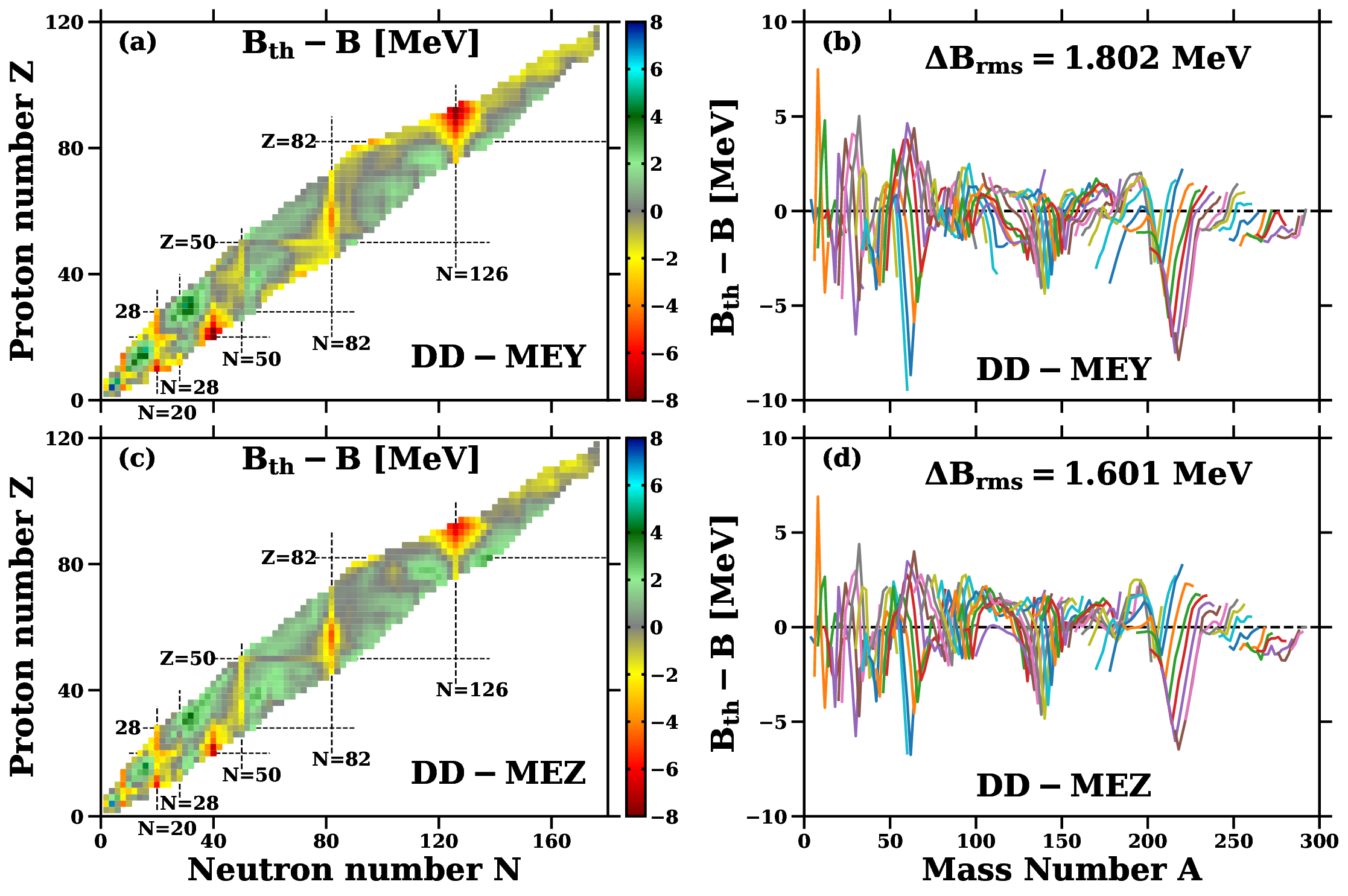}    
    \caption{The differences $B_{th}-B$ between calculated $(B_{th})$ and  experimental 
    nuclear binding energies for indicated DDME functionals (see text for details). 
    \label{compar-DDME}
}
\end{figure*}

   It is important to understand the consequences of the inclusion of new features 
into fitting protocol discussed in Sec.\ \ref{fitting-protocol} on the parameters, 
properties and global  performance of new versions of CEDFs. To do that we 
compare new globally fitted functionals (DD-MEZ, NL5(Z) and PC-Z) (further
Z type of the functionals) with the DD-MEY, NL5(Y) and PC-Y (further Y type of the 
functionals) ones defined in Ref.\ \cite{TA.23}. 

   Note that the Z type of the  functionals has been fitted with accounting of infinite 
basis corrections to binding energies in  fermionic and bosonic sectors and total electron 
binding energies are taken into  account in conversion of atomic binding energies to 
nuclear ones (see Sec.\  \ref{fitting-protocol} and \ref{Pseudo-data}). In contrast these 
corrections are neglected in the fitting protocol of the Y type of CEDFs which was optimized 
in the RHB calculations  carried out in the fermionic and bosonic bases with $N_F=20$ and 
$N_B=20$  (see Ref.\ \cite{TA.23}).  The Z (Y) functionals were fitted to the  AME2020 
\cite{AME2020-second} (AME2016 \cite{AME2016-third}) mass evaluations which contain 
binding energies for 882 (853) even-even nuclei located between proton and neutron drip 
lines.  Note that the same set of 882 even-even nuclei is used in the global comparison 
of nuclear binding energies obtained with the Z and Y types of the functionals.

\subsection{Density dependent meson exchange functionals}
\label{DDME-funct}

    Let us start the comparison with the DD-MEZ and DD-MEY functionals.
Fig.\  \ref{compar-DDME} and Table \ref{table-global-DDME} compare global performance 
of these two functionals.  It substantially improves (by 0.2 MeV) on transition from DD-MEY  
($\Delta B_{rms}=1.802$ MeV) to DD-MEZ  ($\Delta B_{rms} = 1.601$ MeV).
Numerical errors in the calculations of binding energies with these
two functionals are discussed in detail in Sec.\ \ref{calc-error}. The addition of Wigner energy 
(not employed in the fitting protocol) in the formulation of Ref.\ \cite{GSHPT.02} leads to further 
reduction of $\Delta B_{rms}$ down to $\Delta B_{rms} = 1.557$ MeV for the DD-MEZ functional.
The transition from DD-MEY to DD-MEZ leads to $\approx 50$  keV ($\approx 100$ keV) 
reduction in the rms-error for two-neutron (two-proton) separation  energies (see Table 
\ref{table-global-DDME}).  Note that the accuracy of the global  description 
of charge radii is comparable in both functionals (see Table \ref{table-global-DDME}).

   This table also shows that for a given functional the rms-errors in binding energies 
and two-particle separation energies are substantially improved if the light nuclei 
with $A < 70$  are excluded from consideration. This is rather common feature 
seen in many DFT calculations (see, for example, Refs.\ \cite{UNEDF1,AARR.14,YWZL.21}).
There are several possible sources for that. First, on average beyond mean field effects 
neglected in the present paper vary from nucleus to nucleus more drastically in the region 
of light nuclei as compared with the region of medium and heavy mass nuclei (see Fig. 2 in 
Ref.\ \cite{FMXLYM.13}, Fig. 2  in Ref.\ \cite{LLLYM.15} and Fig. 2 in Ref.\ \cite{YWZL.21}).  
Second,  it is also likely that the deficiencies in the description of underlying single-particle 
structure of  very neutron-rich $A<70$ nuclei result in large differences between 
experimental and calculated binding energies seen in Fig.\ \ref{compar-DDME}(a) and (c). 
Third, the deficiencies in isovector channel of the functionals
can reveal themselves in a more pronounced way in light nuclei because of higher isospin
experimentally achievable in these nuclei. To properly identify the contribution of these sources 
to observed features a global fit of the functionals at beyond mean field level is required.

\begin{table}[htb]
\centering
\caption{The parameters of the DDME CEDFs discussed in the present paper. The 
DD-MEY parameters are taken from Supplemental Material of Ref.\ \cite{TA.23}. 
The mass $m$ of the nucleon is ﬁxed at 939 MeV and the mass of the $\omega$-meson 
at 783 MeV. The last column shows the ratio $ratio_i = \frac{par_i(DD-MEZ)}{par_i(DD-MEY)}$ 
of the parameters $par_i$ of the DD-MEZ and DD-MEY functionals.
 }
\centering
\begin{tabular}{cccc}
\hline \hline
Parameter                 &  DD-MEY     & DD-MEZ      & $ratio_i$\\ 
\hline 
$m_\sigma$ [MeV]    & 551.321796  & 558.605889  & 1.013         \\ 
$g_\sigma$                &  10.411867  &  10.602029  & 1.018         \\ 
$g_\omega$               &  12.803298  &  12.881123  & 1.006         \\ 
$g_\rho$                     &   3.692170  &   3.514796  & 0.952         \\ 
$b_\sigma$                 &   2.059712  &   2.667539  & 1.295         \\ 
$c_\sigma$                 &   3.210289  &   4.070271  & 1.268         \\ 
$c_\omega$                &   3.025356  &   4.352258  & 1.439         \\ 
$a_\rho$                      &   0.532267  &   0.649431  & 1.220         \\ 
\hline \hline
\end{tabular} 
\label{table-DDME-par}
\end{table} 

    DD-MEY functional takes total electron binding energies  and infinite basis corrections 
to fermionic and  bosonic bases into account in an effective way (via parameter modifications
during fitting procedure) since they are present in experimental data on atomic binding energies. 
However, explicit accounting of these effects in the DD-MEZ functional leads to quite substantial
(by 22\%$-$44\%) modification of the $b_{\sigma}$, $c_{\sigma}$, $c_{\omega}$ and 
$a_\rho$ parameters which are responsible for explicit density dependence of 
meson-nucleon interaction (see Table \ref{table-DDME-par}). In contrast, the 
modifications of the $m_{\sigma}$, $g_{\sigma}$ and  $g_{\omega}$ parameters are
significantly smaller being on the order of 1\%. These parameters define the scalar
$S$ and vector $V$ potentials in the CDFT which build up the central nucleonic potential.

\begin{table}[htb]
\centering
\caption{The global performance of the DDME CEDFs. The rms deviations
$\Delta B_{rms}$,   $\Delta (S_{2n})_{rms}$, $\Delta (S_{2p})_{rms}$ and $\Delta (r_{ch})_{rms}$ 
between calculated and experimental binding energies $B$, two-neutron (two-proton) 
separation energies $S_{2n}$ ($S_{2p}$), and charge radii $r_{ch}$.  
The  $\Delta (r_{ch})_{rms}$ values are calculated using experimental data on charge radii of 
261 even-even nuclei\footnote{Note that in the analysis of charge radii  we focus on the 
nuclei for which the absolute values of charge radii are experimentally measured and 
for which mean-field approximation is expected to be a reasonably well justified. Thus,
the number of nuclei used in the comparison between theory and experiment is
reduced as compared with that presented in the compilation of Ref.\ \cite{AM.13}
(see Sec. IV.D of Ref.\ \cite{TOAPT.24} for further details).}. The values shown in parentheses 
are the rms deviations for the subset of nuclei which excludes light nuclei with $A < 70$. 
\label{table-global-DDME}
}
\begin{tabular}{ccc}
\hline \hline
Parameter                                   &  DD-MEY             & DD-MEZ          \\  \hline
$\Delta B_{rms}$ [MeV]               & 1.802 (1.506)   & 1.601 (1.416)   \\ 
$\Delta (S_{2n})_{rms}$ [MeV]    & 1.197 (0.751)   & 1.146 (0.782)   \\  
$\Delta (S_{2p})_{rms}$ [MeV]    & 0.951 (0.612)   & 0.847 (0.612)   \\  
$\Delta (r_{ch})_{rms}$ [fm]        & 0.0215 (0.0212) & 0.0254 (0.0261) \\ \hline \hline                 
\end{tabular} 
\end{table} 

\subsection{Nonlinear meson exchange functionals}
\label{NLME-funct}

   The parameters of the NL5(Z) functional are presented in Table \ref{par-NLME}.
One can see that the transition from the NL5(Y) to NL5(Z) functional has very small 
effect on the $m_{\sigma}$, $g_{\sigma}$ and $g_{\omega}$ parameters defining  
central nucleonic potential.  In contrast, the $g_2$ and $g_3$ parameters, which determine 
density dependence of  nonlinear meson self-interaction terms, are substantially
modified by this transition.  This is similar to the situation with  DD-ME functionals discussed in Sec.\ 
\ref{DDME-funct}. Note that the $g_{\rho}$ parameter is only 
weakly modified by this kind of transition both in the NLME  (see Table \ref{par-NLME}) 
and DDME  (see Table \ref{table-DDME-par}) functionals.

\begin{table}[htb]
\begin{center}
\caption{The parameters of the NLME CEDFs discussed in the present paper. 
              The NL5(Y) parameters are taken from Ref.\ \cite{TA.23}. The mass 
              $m$ of the nucleon is fixed at 939 MeV, the  mass of $\omega$-meson 
              at 782.6 MeV and the mass of $\rho$-meson at 763 MeV. The last column 
              shows the ratio $ratio_i$ of the parameters $par_i$ of the NL5(Z) 
              and NL5(Y) functionals.                            
\label{par-NLME}
}
\begin{tabular}{cccc} \hline \hline
                          &   NL5(Y)     & NL5(Z)       & $ratio_i$  \\ \hline
 $m_{\sigma}$  [MeV]      & 508.628825   & 515.823458   & 1.014          \\ 
 $g_{\sigma}$             &  10.274835   &  10.359299   & 1.008          \\ 
 $g_{\omega}$             &  12.906821   &  12.920545   & 1.001          \\ 
 $g_{\rho}$               &   4.375307   &   4.382018   & 1.002          \\ 
 $g_{2}$      [fm$^{-1}$] & $-11.197664$ & $-9.786977$  & 0.874          \\ 
 $g_{3}$                  & $-31.996125$ & $-26.395135$ & 0.825          \\ \hline \hline 
\end{tabular}
\end{center}
\end{table}

\begin{figure}[htb]
\centering
\includegraphics[width=8.5cm]{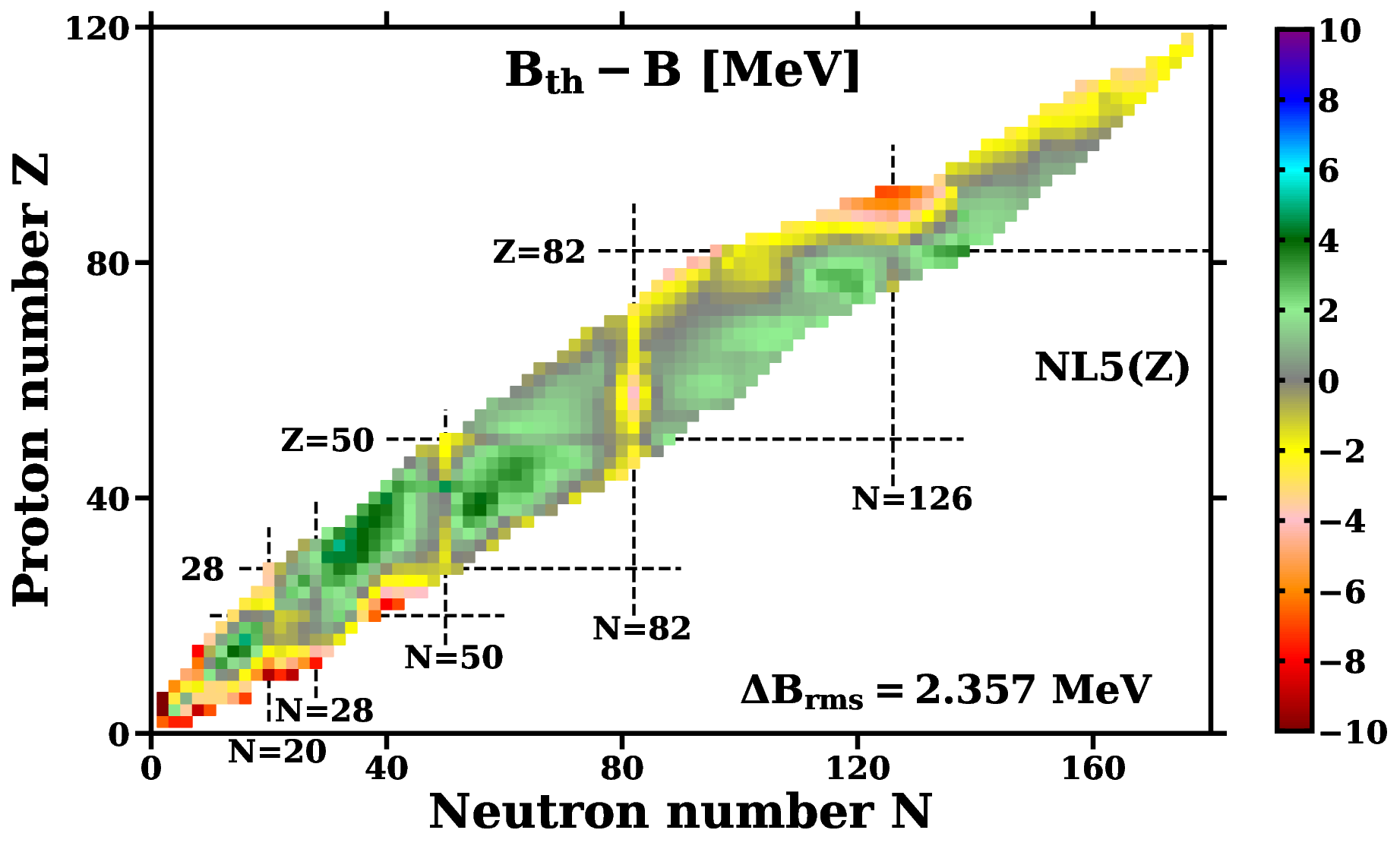}
    \caption{The differences $B_{th}-B$ between calculated and experimental
     nuclear binding energies for the NL5(Z) functional.
         \label{compar-NL5Z}
    }
\end{figure}

The global performance of the NL5(Z) functional in the description of the 
ground state properties is presented in Fig.\ \ref{compar-NL5Z} and Table 
\ref{table-global-NLME}.  The NL5(Z) and NL5(Y) functionals perform comparably: 
there is only small improvement in global description of physical observables of 
interest on the transition from NL5(Y) to NL5(Z).

\begin{table}[htb]
\caption{The same as Table \ref{table-global-DDME} but for the NLME CEDFs.
}
\centering
\begin{tabular}{ccc}
\hline \hline
Parameter                      &  NL5(Y)         &  NL5(Z)        \\ \hline
$\Delta B_{rms}$ [MeV]         & 2.407 (1.768)   & 2.357 (1.914)   \\ 
$\Delta (S_{2n})_{rms}$ [MeV]  & 1.253 (0.711)   & 1.238 (0.710)   \\  
$\Delta (S_{2p})_{rms}$ [MeV]  & 1.239 (0.773)   & 1.220 (0.816)   \\  
$\Delta (r_{ch})_{rms}$ [fm]   & 0.0277 (0.0276) & 0.0274 (0.0270) \\ \hline \hline                        
\end{tabular} 
\label{table-global-NLME}
\end{table} 

\subsection{Point coupling functionals}
\label{PC-funct}

\begin{figure}[htb]
	\centering
        \includegraphics[width=8.5cm]{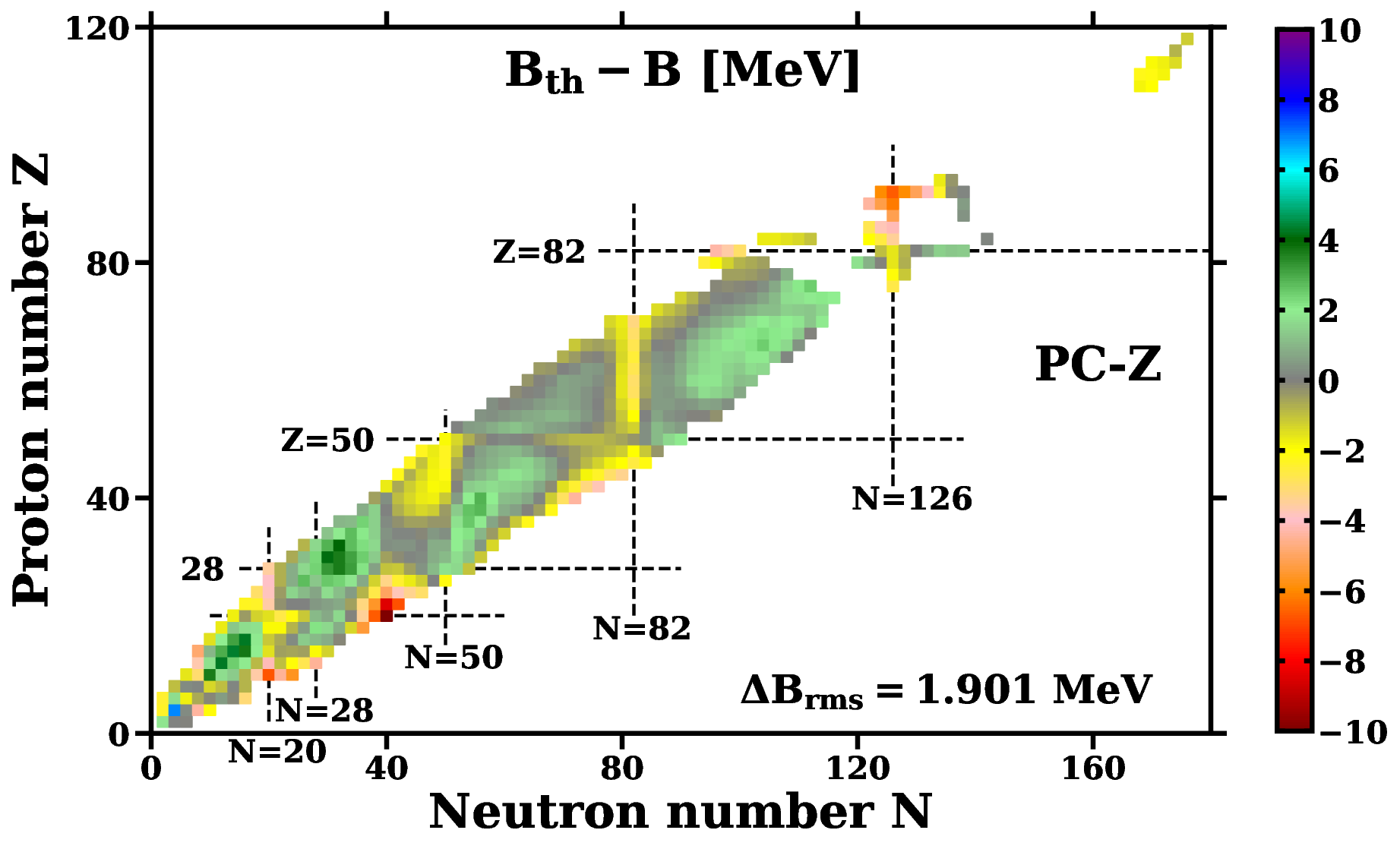}
\caption{The same as Fig.\ \ref{compar-NL5Z} but for the PC-Z functional.
  Only the nuclei for which infinite basis corrections to the binding energies
  can be defined in the fermionic basis are shown here.     
\label{compar-PC-Z}
}
\end{figure}

   The results presented in Fig.\ \ref{290-LV-and-others} of the present paper  
and in Sec. VI of Ref.\ \cite{TOAPT.24}  show that the PC functionals are 
characterized by the worst convergence of binding energies as a function of 
$N_F$ amongst considered classes of CEDFs. As a result, for the PC-Z functional 
the infinite basis corrections to the binding energies due to truncation of fermionic 
basis can be defined only for the sublead region and for some nuclei which have 
spherical shape in lead and the superheavy region (see Fig.\ \ref{compar-PC-Z}). 
Thus, the fitting  protocol of PC-Z is biased 
as compared with the one for the DD-MEZ and NL5(Z) functionals by exclusion 
of the majority of the nuclei in the lead region, actinides and superheavy nuclei 
from the definition of correction function given by Eq. (1) in Ref.\ \cite{TA.23}. The 
PC-Z  functional is characterized by  the rms error of $\Delta B_{rms}= 1.901$ 
MeV  in the description of nuclear binding energies (see Fig.\ \ref{compar-PC-Z}).

     Because of above mentioned limitations we compare only the parameters of 
the  PC-Z and PC-Y1 functionals which are presented in Table \ref{par-PC-Z}:
no comparisons of global performance and nuclear matter properties of these
two functionals are presented here. Table  \ref{par-PC-Z} shows that the $\alpha_S$, 
$\alpha_V$, and $\alpha_{TV}$ parameters are very similar in those two functionals 
and slightly larger difference exists for the $\beta_S$ one.  In contrast, the differences 
between two functionals are large for the $\gamma_S$, $\gamma_V$, $\delta_S$,
$\delta_V$ and $\delta_{TV}$ parameters (see last column in Table \ref{par-PC-Z}).
The  first two parameters (as well as $\beta_S$) are responsible for medium dependence 
of the effective interaction, while the latter three simulate the effects of finite range (see 
Ref.\ \cite{PC-PK1}). Note that the $\gamma_S$ and $\gamma_V$ parameters  as well 
as the $\delta_S$  and $\delta_V$ ones are parametrically correlated in the PC class of 
the  functionals with substantial range of permissible variations of the parameters (see 
Fig. 4 in Ref.\ \cite{TAAR.20} and its  discussion).

\begin{table}[htb]
\centering
\caption{The parameters of the PC CEDFs discussed in the present paper.  The 
PC-Y1  parameters are taken from Supplemental Material of Ref.\ \cite{TA.23}.
The last column shows the ratio $ratio_i$ of the parameters $par_i$ of the PC-Z 
and PC-Y1 functionals.
}
\begin{tabular}{cccc}
\hline \hline
                            &        PC-Y1         &      PC-Z      &      $ratio_i$ \\ \hline
$\alpha_S$    [MeV$^{-2}$]  &   $-0.394920^{-03}$  &   $-0.395616^{-03}$  & 1.002      \\ 
$\alpha_V$    [MeV$^{-2}$]  &   $ 0.268670^{-03}$  &   $ 0.270967^{-03}$  & 1.009      \\ 
$\alpha_{TV}$ [MeV$^{-2}$]  &   $ 0.293450^{-04}$  &   $ 0.287055^{-04}$  & 0.978      \\ 
$\beta_S$     [MeV$^{-5}$]  &   $ 0.832870^{-10}$  &   $ 0.771989^{-10}$  & 0.927      \\
$\gamma_S$    [MeV$^{-8}$]  &   $-0.344690^{-16}$  &   $-0.257634^{-16}$  & 0.747      \\
$\gamma_V$    [MeV$^{-8}$]  &   $-0.463780^{-17}$  &   $-0.805090^{-17}$  & 1.736      \\ 
$\delta_S$    [MeV$^{-4}$]  &   $-0.212860^{-09}$  &   $-0.830206^{-10}$  & 0.390      \\ 
$\delta_V$    [MeV$^{-4}$]  &   $-0.336750^{-09}$  &   $-0.443813^{-09}$  & 1.318      \\ 
$\delta_{TV}$ [MeV$^{-4}$]  &   $-0.470160^{-09}$  &   $-0.573487^{-09}$  & 1.220      \\ \hline \hline
\end{tabular}
\label{par-PC-Z}
\end{table}

\subsection{On the connection between global binding energies of the ground 
                    states and nuclear matter properties}
\label{NMP}
  
   An important question is {\it whether nuclear matter properties (NMPs) can be 
defined from the global analysis of binding energies.} Externally defined NMPs 
(see, for example, Refs.\ \cite{Sk-nm,RMF-nm})  are used as additional constraints in 
existing global fits of the parameters to binding energies  in some versions of the 
microscopic+macroscopic (mic+mac) model (see Ref.\ \cite{MSIS.16}) and in 
non-relativistic DFTs (see, for example, Refs.\ \cite{GCP.10,D1M*}). However, the 
FRDLM version of the mic+mac model defines NMPs from mass-like quantities (see 
Ref.\ \cite{MSIS.16}).

  The fitting protocols of absolute majority of CEDFs employ externally defined NMPs as 
constraints. However,  this is not always the case. For example, the PC-PK1 CEDF has 
been defined by the fit to only binding energies and charge radii of 60 nuclei without any 
information on NMPs (see Ref.\ \cite{PC-PK1}).  Nevertheless, obtained values of  the 
symmetry energy $J=35.6$ MeV and its slope $L_0=113$ MeV deviate from those recommended 
for covariant EDFs in Ref.\ \cite{RMF-nm} (see also second column in Tables \ref{table-NMP-NLME} 
and \ref{table-NMP-DDME}). In addition, the DD-MEY  functional globally fitted to binding 
energies describes well NMPs without their inclusion into fitting protocol (see Ref.\ \cite{TA.23} 
and Table \ref{table-NMP-DDME}).  Note that the inclusion of the NMP constraints into the fitting 
protocol does not guarantee that  obtained $K_0$, $J$ and $L_0$  values are located within the 
SET2b ranges.

  Tables \ref{table-NMP-NLME} and \ref{table-NMP-DDME} illustrate the impact of 
the transition from the Y to Z types of the functionals on the NMPs for the DDME and 
NLME classes of CEDFs. This impact is best illustrated by the ratio of a given NMP 
obtained with the Z and Y classes of the functionals provided in the last column of
these two tables. One can see that the $E/A$ and $\rho_0$  (as well as $J$ and 
$L_0$ in the case of the NLME functionals) values are only slightly (within approximately 
2\% of  absolute value)  affected by this transition. The impact of  this transition on 
the $J$ value is somewhat more pronounced (by $\approx 5$\%) in the case of the 
DDME functionals (see Table \ref{table-NMP-DDME}). This transition causes quite 
large change (by $\approx 15$\%) of the slope of the symmetry energy $L_0$ in the 
case  of the DDME functionals.  However, consistently (among considered types of the 
functionals) large impact of  this transition is seen for the incompressibility $K_0$ 
which is increased by $\approx 7$\% and $\approx 12$\% for the DDME and NLME 
functionals,  respectively.  As a result of this increase the $K_0$ values 
calculated with the DD-MEZ and NL5(Z) functionals are located above the SET2b 
upper limit for $K_0$.

    These modifications of NMPs clearly illustrate the importance of taking into 
account the infinite basis  corrections and total electron binding energies into the 
fitting protocol of  CEDFs.

\begin{table}[htb]
\centering
\caption{Symmetric nuclear matter properties of the NLME CEDFs at saturation.
The binding energy per particle $E/A$, the density $\rho_0$, the incompressibility 
$K_0$, the symmetry energy $J$, and the slope of the symmetry energy $L_0$ of 
the functionals under study are shown.  The second column shows 
the SET2b  limits for experimental/empirical ranges for these quantities  (see Ref.\ 
\cite{RMF-nm}). The last column shows the 
$ratio_j = \frac{{\rm NMP}_j{\rm (NL5(Z))}} {{\rm NMP}_j{\rm (NL5(Y))}}$ of 
the $j$-th NMP obtained with the NL5(Z) and NL5(Y) functionals. 
}
\begin{tabular}{ccccc}
\hline \hline
                                    & SET2b            &  NL5(Y)    &   NL5(Z)  &  $ratio_j$ \\  \hline
$E/A$ [MeV]               & $\approx -16$  &  -16.1       &  -16.1      &   1.0         \\ 
$\rho_0$ [fm$^{-3}$]   & $\approx 0.15$  & 0.148       &  0.146     &    0.986    \\ 
$K_0$ [MeV]               &    $190-270$    & 254.5       &  290.4     &   1.141     \\ 
$J$ [MeV]                    &    $25-35$        & 36.6         &   36.1      &   0.986     \\ 
$L_0$ [MeV]                &    $30-80$        & 116.7       &  114.1     &   0.978     \\ 
\hline \hline
\end{tabular} 
\label{table-NMP-NLME}
\end{table} 
   
\begin{table}[htb]
\centering
\caption{The same as Table \ref{table-NMP-NLME} but for DDME CEDFs.}
\begin{tabular}{ccccc}
\hline \hline
                                    & SET2b               &  DD-MEY  & DD-MEZ  &  $ratio_j$         \\  \hline
$E/A$ [MeV]                & $\approx -16$    &  -16.1        & -16.0        &   0.994            \\ 
$\rho_0$ [fm$^{-3}$]   &  $\approx 0.15$  &  0.153        &  0.150      &   0.980             \\ 
$K_0$ [MeV]               &  $190-270$         & 265.8        & 286.6       &  1.078              \\ 
$J$ [MeV]                   &   $25-35$             & 32.8          &  31.3        &  0.954              \\ 
$L_0$ [MeV]               &  $30-80$              & 51.8          &  44.1        &  0.851              \\ 
\hline \hline
\end{tabular} 
\label{table-NMP-DDME}
\end{table} 

\subsection{Global calculation errors due to approximations in the fitting protocols.}
\label{calc-error}

    One interesting question emerging from the present study is what are  the 
consequences of approximations employed in the fitting protocols on different physical  
observables. The previous generations of CEDFs were fitted in finite fermionic and 
bosonic basis  (in the best cases with $N_F=20$ and $N_B=20$ (see, for example, 
Refs.\ \cite{NL3*,PC-PK1,DD-ME2,TA.23} and discussion in Sec.\ V of Ref.\ 
\cite{TOAPT.24}) and with neglect of total electron binding energies in conversion 
of experimental binding energies from atomic to nuclear ones (see Sec.\ \ref{electron}). 
As compared with the results obtained in the Z class of functionals, these approximations 
lead to relatively smooth deviation trends seen in Figs.\ \ref{corr-energy-new} (due to 
neglect of total electron binding energies),  \ref{corr-energy-ferm-global} and 
\ref{corr-energy-bos-global} (due to use of finite $N_F=20$ fermionic and $N_B=20$ 
bosonic bases as compared with infinite ones). The EDF fits based on such approximations 
assume that these smooth deviation trends are accounted for via the modifications of the 
model parameters but the accuracy of such assumptions was never verified before in 
the CDFT.

\begin{figure}[htb]
	\centering
        \includegraphics[width=8.5cm]{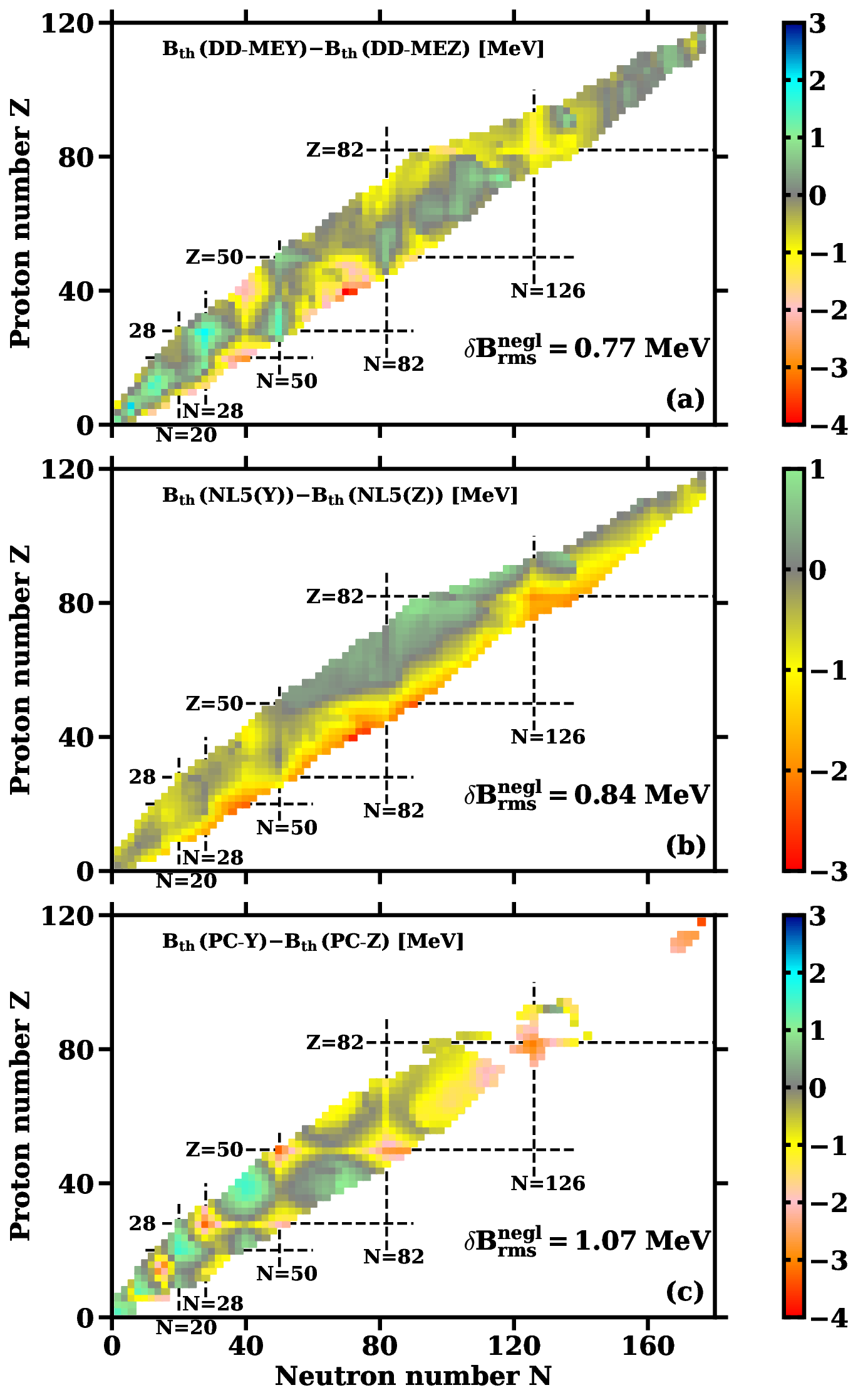}       
\caption{The differences between nuclear binding energies calculated with
the Y and Z classes of the functionals.
\label{compar-exact-approx}
}
\end{figure}

    The accuracy of the description of experimental nuclear binding energies 
in model calculations is described by the rms deviation
\begin{eqnarray}
\Delta B_{rms}^{tot} = \Delta B_{rms} \pm \delta B_{rms}^{negl} 
\end{eqnarray}    
  where  $\Delta B_{rms}$ is the rms deviation as defined from the difference of 
calculated  and  experimental nuclear binding energies and $\delta B_{rms}^{negl}$ is the 
global calculation error due to the truncation of the basis and neglected terms (such as 
total electron binding energy in the conversion from experimental atomic to nuclear 
binding energies).  
The $\delta B_{rms}^{negl}$ error emerges due to the convergence of the optimization
to a different minimum in the parameter hyperspace when above mentioned factors are
neglected  as compared with the one obtained when they are taken into account (see Tables 
\ref{table-DDME-par} and \ref{par-NLME}).
The definition of $\delta B_{rms}^{negl}$ requires the calculations 
in very large bases accurately approximating infinite bases (or reliable evaluation of infinite 
basis correction to binding energies). Existing global mass calculations in the CDFT presented 
in Refs.\ \cite{LRR.99,GTM.05,RA.11,AARR.14,ZNLYM.14,LLLYM.15,AGC.16,YWZL.21,TA.23} 
are carried out in finite fermionic and bosonic bases. As a consequence, $\delta B_{rms}^{negl}$ 
was not defined for them in original publications.

   The present study allows for the first time to evaluate $\delta B_{rms}^{negl}$ in 
global mass calculations within the CDFT framework.  For  example, $\delta B_{rms}^{negl} \approx 28$ 
keV for the DD-MEZ functional.
This error is due to combination of numerical errors in the definition of binding energies corresponding 
to infinite basis in the fermionic and bosonic sectors of CDFT and the $\varepsilon$ value used in the 
termination of the ABOA rounds (see Sec.\ \ref{Pseudo-data} for details) and it is dominated by the errors 
in the fermionic sector. In  contrast, the DD-MEY functional is fitted in finite $N_F=20$ and $N_B=20$ 
bases with neglect of total electron binding energies. Note that fitting protocols of these two functionals 
are almost the same: minor  difference in the number of even-even nuclei used does not play a principal 
role.  Fig.\ \ref{compar-exact-approx}(a) shows that the neglect of above mentioned ingredients in the 
fitting protocol of DD-MEY leads to the binding energies which locally deviate (by up to several MeV) from 
more exact ones obtained with DD-MEZ. These  deviations are caused to a large degree  by the differences 
in the single-particle structure between these two  functionals emerging from the differences in the parameters 
of the functionals (see Table \ref{table-DDME-par}). These differences in single-particle structure affect shell 
correction contributions to binding energies. Fig.\ \ref{compar-exact-approx}(a) allows to define 
$\delta B_{rms}^{negl}=0.77$ MeV for  the DD-MEY functional using the DD-MEZ results as a benchmark 
(see  Fig.\ \ref{compar-exact-approx}(a)).

    The same situation is seen when comparing the NL5(Y) and NL5(Z) functionals, 
see  Fig.\ \ref{compar-exact-approx}(b)). This figure gives $\delta B_{rms}^{negl}=0.84$ MeV for 
the NL5(Y) functional with respect of the NL5(Z) benchmark. Note that $\delta B_{rms}^{negl}\approx 0.023$ 
MeV for NL5(Z) functional. Moreover, the neglect of infinite basis corrections and total  electron binding 
energies in the fitting protocol of  the NL5(Y) functional leads not only to local fluctuations in 
B$_{\rm th}$(NL5(Y))$-$B$_{\rm th}$(NL5(Z)) but  also to the appearance of wrong isospin trends 
in the binding energies as compared with the NL5(Z) functional.

   Let us focus on the DDME class of the functionals. Based on previous discussion 
one gets $ \Delta B_{rms}^{tot}(${\rm DD-MEZ}$) = 1.601  \pm 0.028$ MeV 
[or $ \Delta B_{rms}^{tot}(${\rm DD-MEZ}$) = 1.557  \pm 0.028$ MeV when Wigner energy is
included]  and   $\Delta B_{rms}^{tot}(${\rm DD-MEY}$) = 1.802  \pm 0.77$ MeV for  the DD-MEZ and
DD-MEY functionals, respectively. $\Delta B_{rms}(${\rm DD-MEZ}$)=1.601$ MeV 
is by approximately 200 keV better than $\Delta B_{rms}(${\rm DD-MEY}$)=1.802$ MeV.
However, it is even more important that the DD-MEZ functional is characterized by drastically 
improved global calculation error as compared with DD-MEY.  This error 
($\delta B_{rms}^{negl}(\rm DD-MEY)=0.77$ MeV) is too large ($\approx 43\%$ of 
$\Delta B_{rms}$) for  DD-MEY functional.

   Other existing CDFT global mass calculations do not include the assessment 
of global calculation errors due to the use of finite fermionic and bosonic bases and neglect 
of total electron binding energies. However, considering the similarity of their fitting approaches 
and global calculations in respect of these approximations to that of the Y class of the functionals, 
it is reasonable to expect that $\delta B_{rms}^{negl}$ is of the order of 0.8 MeV for 
the CDFT mass tables based on the DDME and NLME functionals presented in Refs.\ 
\cite{LRR.99,GTM.05,RA.11,AARR.14}. 

  In contrast, at present it is impossible to accurately quantify the $\delta B_{rms}^{negl}$ values 
for the global mass tables  obtained with the PC functionals  (such as PC-PK1 (see Refs.\ 
\cite{ZNLYM.14,LLLYM.15,YWZL.21}),  PC-Y (see Ref.\ \cite{TA.23}), and DD-PC1 
(see Ref.\ \cite{AARR.14}). This is due to the difficulties of obtaining the binding 
energies corresponding to infinite fermionic basis in actinides and superheavy 
nuclei discussed  for this class of functionals in Sec.\ \ref{Pseudo-data}. In addition, the 
$\Delta B^{cor}(Z,N)$ values (see Eq.\ \ref{correct-func}) are the largest among considered 
classes of the functionals for the PC ones since the contributions to this quantity due to 
total electron binding energies and infinite basis corrections in the fermionic sector act in 
the same direction. Both these factors suggest that the $\delta B_{rms}^{negl}$ values 
should be larger than those for the NLME and DDME functionals. Indeed the  comparison 
of the PC-Y and PC-Z functionals suggest that  $\delta B_{rms}^{negl}$ is larger than 
1.0 MeV (see Fig.\ \ref{compar-exact-approx}(c))\footnote{In Refs.\ 
\cite{ZNLYM.14,LLLYM.15,YWZL.21}, the harmonic oscillator basis with $N_F'=12$, 
14 and 16 fermionic shells is used for the calculations of the nuclei with $Z<20$, 
$20\leq Z <82$ and $82\leq Z \leq 104$,  respectively. However, the PC-PK1 functional
used in these calculations has been fitted in Ref.\ \cite{PC-PK1}  in truncated basis 
with $N_F=20$. The RHB calculations comparing the binding energies obtained with
$N_F'$ and $N_F=20$ clearly show that this difference in the basis size contributes
itself additional 0.199 MeV to above mentioned value of  $\delta B_{rms}^{negl}[{\rm PC}]\approx 1.0$ 
MeV. Thus, this value $\delta B_{rms}^{negl}$ is very likely the conservative (lowest) 
estimate for global calculation errors in the case of the PC functionals. 
}.

   Above discussed global calculation errors of the order of 0.8 MeV are 
unacceptable  if one aims at the accuracy of the global  description of binding energy  at the level of 
$\Delta B_{rms} \approx 0.5$ and 0.8 MeV which is achieved in global mass fits in non-relativistic 
Skyrme and Gogny DFTs, respectively (see Refs.\  \cite{GCP.13,GHGP.09}). Note that such calculations 
always include infinite basis corrections to binding energies (see Refs.\ \cite{HG.07,GHGP.09,BCPM}). 
This clearly indicates the need for use of infinite basis corrections to binding energies and accounting 
of total electron binding energies in high-precision global binding energy fits within the CDFT.

   Note that charge radii and separation energies are also affected by global calculation 
errors.  These errors are  very small for the Z class of functionals: $\delta (r_{ch})_{rms}^{negl} \leq 0.0001$ 
fm for charge radii and $\delta (S_{2p})_{rms}^{negl} \leq 2$ keV, $\delta (S_{2n})_{rms}^{negl} \leq 2$ 
keV for two-particle separation energies.  In contrast, they are quite substantial for the Y class of the 
functionals. For example, $\delta (r_{ch})_{rms}^{negl} = 0.010$ fm and 0.009 fm for the DD-MEY and 
NL5(Y) functionals with respect of accurate benchmarks provided by the DD-MEZ and NL5(Z) 
functionals. Based on these results and those presented in Tables 
\ref{table-global-DDME}  and \ref{table-global-NLME},  $\Delta (r_{ch})_{rms}^{tot} = 0.0215 \pm 0.01$ 
fm and $\Delta (r_{ch})_{rms}^{tot} = 0.0277 \pm 0.008$ fm for the DD-MEY and NL5(Y) functionals,
respectively. As a result, these global calculation errors have to be taken into account when the charge 
radii of the Y and Z functionals are compared.

   Global calculation errors for two-particle separation energies in the Y class of the
functionals are of the order of  0.3 MeV with respect of the benchmarks provided by the Z class of
the functionals.  More precisely  $\Delta (S_{2n})_{rms}^{tot} = 
1.197 \pm 0.387$ MeV  ($\Delta (S_{2n})_{rms}^{tot} = 0.751 \pm 0.343$ MeV) and 
$\Delta (S_{2p})_{rms}^{tot} = 0.951 \pm 0.385$ MeV  ($\Delta (S_{2p})_{rms}^{tot} = 
0.612 \pm 0.334$ MeV) for DD-MEY CEDF. Note that the values of $\Delta (S_{2n})_{rms}$ 
and $\Delta (S_{2p})_{rms}$ are taken from Table \ref{table-global-DDME} and the values in 
parentheses correspond to the subset in which the $A<70$ nuclei are excluded. In a similar 
fashion  $\Delta (S_{2n})_{rms}^{tot} = 1.253 \pm 0.211$  MeV  
($\Delta (S_{2n})_{rms}^{tot} = 0.711 \pm 0.205$ MeV) and  
$\Delta (S_{2p})_{rms}^{tot} = 1.239 \pm 0.261$ MeV  
($\Delta (S_{2p})_{rms}^{tot} = 0.773 \pm 0.263$) MeV for the NL5(Y)
functional.

\section{Conclusions}
\label{Concl}

   The present study represents a further development of covariant energy density
functionals towards more accurate description of binding energies across the
nuclear chart. The following results have been obtained:

\begin{itemize}
\item
     For the first time, infinite basis corrections to binding energies due to fermionic 
and mesonic bases have been taken into account  in the fitting protocol of CEDFs 
in the CDFT framework. They eliminate  numerical uncertainties related to the use 
of finite bases which exist in global fits of relativistic (see Refs.\ \cite{AGC.16,TA.23}) 
and  non-relativistic (see, for example, Refs.\ \cite{UNEDF0,UNEDF2}) functionals in 
which the basis with finite number of HO shells (typically, with less than  or equal to
$N_F=20$ HO shells) is used.

\item
   For the first time, total electron binding energies $B_{el}$ have been 
taken into account in the conversion of experimental atomic binding energies presented 
in AME compilations into nuclear ones in the fitting protocol of CEDFs. This step has 
either been omitted or overlooked in the fitting protocols of many EDFs (see Sec.\ 
\ref{electron}).   State-of-the-art atomic calculations reveal that the increase of
nuclear charge radii with increasing neutron number has very small impact on 
$B_{el}$. As a consequence, the values of total electron binding energies provided
in Ref.\ \cite{DFA.24} can be safely used for atoms of all nuclei in experimentally known
nuclear chart.

\item
   The accuracy of global description of binding energies is defined not only 
by rms deviation between calculated and experimental nuclear binding energies
but also by global calculation error $\delta B_{rms}^{negl}$. The latter is not defined
in previous global mass calculations. In existing global calculations within the 
CDFT, it  emerges from the truncation of fermionic and bosonic bases and the neglect 
of total  electron binding energies in conversion of experimental atomic binding into 
nuclear ones. For the first time, it was illustrated on the example of DD-MEY and 
NL5(Y) functionals that these global calculation errors are non-negligible being on the 
order of $\delta B_{rms}^{negl} \approx 0.8$ MeV. The accounting of these factors 
reduced $\delta B_{rms}^{negl}$ to approximately 0.025 MeV in the DD-MEZ and NL5(Z) 
functionals. Thus, infinite basis corrections to binding energies and total electron 
binding energies have to be taken into account in the global fits of CEDFs if one 
aims at the accuracy  of $\Delta B_{rms} \approx 0.5$ and 0.8 MeV achieved in 
non-relativistic Skyrme  and Gogny DFTs, respectively (see Refs.\  
\cite{GCP.13,GHGP.09}).

\item
   The DD-MEZ
functional provides the best global accuracy of the description of binding energies (i.e.  
$\Delta B_{rms}^{tot}(${\rm DD-MEZ}$) = 1.557  \pm 0.025$ MeV [with Wigner energy
included] for 882 even-even nuclei) among existing CEDFs at the mean field level. 
This value is comparable with  $\Delta B_{rms} = 1.518$ MeV obtained for a smaller 
set of 637 nuclei (see Ref.\ \cite{RHB-e-e-continuum.22}) in the 
RHB calculations in continuum with the PC-PK1 functional and phenomenological 
treatment of beyond mean field effects\footnote{In this paper, the Dirac 
Woods-Saxon basis is used and the numerical errors in the calculations of binding 
energies are substantially reduced as compared with those of the global calculations 
presented in Refs.\ \cite{ZNLYM.14,LLLYM.15,YWZL.21} which also employ the PC-PK1 
functional but are carried out in harmonic oscillator basis.
As a consequence, the results presented in these references suffer from  slow convergence of 
binding energies as a function of $N_F$ (see discussion in Sec.\ \ref{Pseudo-data} of the 
present paper and Sec.\ VI of Ref.\ \cite{TOAPT.24}). However, similar to Refs.\ 
\cite{ZNLYM.14,LLLYM.15,YWZL.21}  the results presented in Ref.\ \cite{RHB-e-e-continuum.22} 
are based on approach which still neglects total  electron binding energies leading to appreciable 
global  calculation errors in binding energies.}. 
Further substantial improvement of global accuracy of the description of binding energies in
the DDME functionals is expected when beyond mean field effects are included into global 
fitting protocol since these effects typically improve $\Delta B_{rms}$ by approximately
1 MeV (see, for example,  Table A in Ref.\  \cite{RHB-e-e-continuum.22}).

\end{itemize}
 
\section{ACKNOWLEDGMENTS}

   This material is based upon work supported by the U.S. Department of Energy,  
Office of Science, Office of Nuclear Physics under Award No. DE-SC0013037
and by the Australian Research Council Grants No. DP230101058 and 
DP200100150.

\bibliography{references-45-masses-BMF.bib}

\newpage

.

\newpage


\centerline{\bf \Large Supplemental material}









\setcounter{figure}{0} 
\setcounter{table}{0}
\setcounter{section}{0}

\section{Overview}
    Supplemental material provides additional information to the manuscript. Field isotopic 
shifts to total electron binding energies of the atoms of superheavy nuclei are provided in 
Sec.\ \ref{field}. Sec.\ \ref{iter-con} describes and illustrates the iterative procedure used 
for accurate definition of infinite basis corrections to binding energies in fermionic  basis.

\section{Field isotopic shifts to total electron binding energies of the atoms of superheavy
nuclei.} 
\label{field}

   Table \ref{t:etot}  displays field isotope shift constants $F$ of the atoms of the Pb nucleus 
and superheavy elements with $Z=100-120$. Note that the $F$ values drastically
decline with decreasing $Z$ so this is a reason why they are not shown for $Z\leq 82$.
Note that additional information on total electron binding  energies of these configurations 
are provided in Table III of Ref.\ \cite{DFA.24}.

\section{The iterative procedure for definition of infinite basis corrections to binding
energies in fermionic basis.
}
\label{iter-con}

\begin{table} 
\caption{\label{t:etot}  
Field isotope shift constants $F$ of the atoms of the Pb nucleus and superheavy 
elements with $Z=100-120$ (column 5).  The columns 3 and  4 show the ground 
state atomic configurations and their total electron angular momentum $J$. }
\begin{ruledtabular}
\begin{tabular}   {ccccc}
\multicolumn{1}{c}{$Z$}&
\multicolumn{1}{c}{Atom}&
\multicolumn{2}{c}{Ground state}&
\multicolumn{1}{c}{$F$}\\
&&\multicolumn{1}{c}{config.}&
\multicolumn{1}{c}{$J$}&
\multicolumn{1}{c}{(keV/fm$^2$)}\\
\hline
1  &   2  &           3                   &  4   &    5       \\
 82 & Pb & [Hg]$6p^2$        &  0   & 0.0041 \\
100 & Fm & [Rn]$5f^{12}7s^2$ &  6   & 0.022  \\
101 & Md & [Rn]$5f^{13}7s^2$ & 7/2  & 0.024  \\
102 & No & [Rn]$5f^{14}7s^2$ &  0   & 0.026 \\
103 & Lr & [No]$6d$          & 3/2  & 0.029 \\
104 & Rf & [No]$6d^2$        &  2   & 0.032 \\
105 & Db & [No]$6d^3$        & 3/2  & 0.035 \\
106 & Sg & [No]$6d^4$        &  0   & 0.038 \\
107 & Bh & [No]$6d^5$        & 5/2  & 0.042 \\
108 & Hs & [No]$6d^6$        &  4   & 0.046 \\
109 & Mt & [No]$6d^7$        & 9/2  & 0.055 \\
110 & Ds & [No]$6d^8$        &  4   & 0.056 \\
111 & Rg & [No]$6d^9$        & 5/2  & 0.062 \\
112 & Cn & [No]$6d^{10}$     &  0   & 0.068 \\
113 & Nh & [Cn]$7p$          & 1/2  & 0.075 \\
114 & Fl & [Cn]$7p^2$        &  0   & 0.082 \\
115 & Mc & [Cn]$7p^3$        & 3/2  & 0.091 \\
116 & Lv & [Cn]$7p^4$        &  2   & 0.100 \\
117 & Ts & [Cn]$7p^5$        & 3/2  & 0.111 \\
118 & Og & [Cn]$7p^6$        &  0   & 0.118 \\
119 & E119 & [Og]$8s$        & 1/2  & 0.135 \\
120 & E120 & [Og]$8s^2$      &  0   & 0.149 \\  
\end{tabular}
\end{ruledtabular}
\end{table}

\begin{figure}[htb]
  \centering
\includegraphics[width=8.0cm]{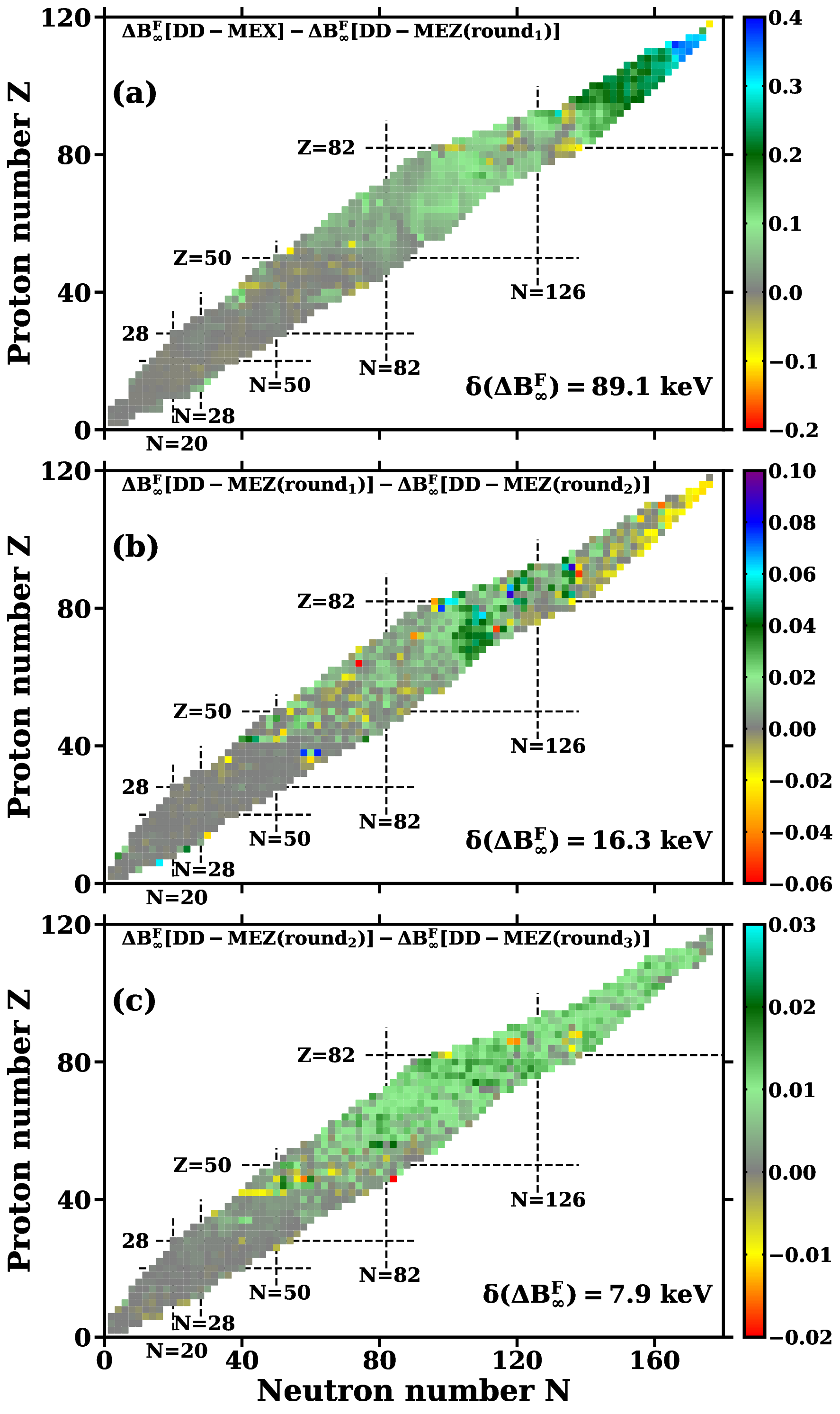}
\caption{ The illustration of the iterative procedure for definition of infinite basis corrections 
to binding energies in fermionic basis on the example of the DD-MEZ funcitonal. The colormap 
shows the difference of infinite basis corrections to binding energies [in MeV] obtained with 
two indicated versions of the functionals. Note that the energy range of the colormap depends
on panel.  See text for details.
\label{iter-proc} 
}
\end{figure}

\begin{figure}[htb]
  \centering
\includegraphics[width=8.0cm]{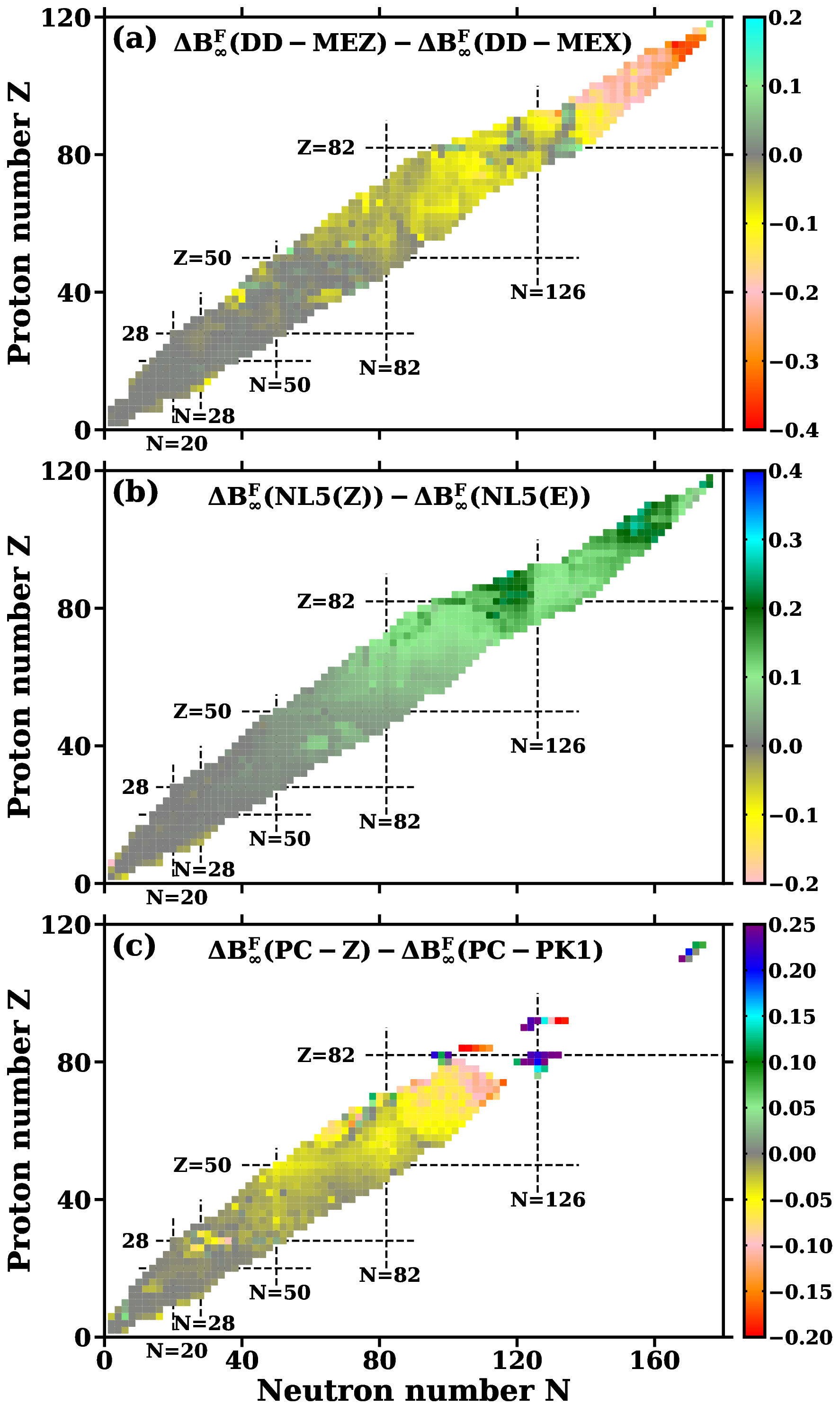}
\caption{The difference [in MeV] of infinite basis corrections to binding energies in the 
              fermionic basis obtained with indicated pairs of the functionals.
\label{iter-difference} 
}
\end{figure}

   Fig. \ref{iter-proc} illustrates iterative procedure for the definition of infinite basis 
corrections to binding energies in  fermionic basis.  The initial ABOA round of the optimization 
of the DD-MEZ  functional is done employing infinite basis corrections obtained for the DD-MEX 
functional (see Fig.\ 10(a) in Ref.\ \cite{TOAPT.24}). It leads to the DD-MEZ(round$_1$) version 
of the functional. Then infinite basis corrections  $\Delta B_{\infty}^{F}(Z,N)$[DD-MEZ(round$_1)$] 
are defined using the procedure described in Sec. VI of Ref.\ \cite{TOAPT.24}.
Fig.\ \ref{iter-proc}(a) shows that the difference between infinite basis corrections 
obtained with these two functionals is quite substantial in medium and heavy mass nuclei approaching 
400 keV  in superheavy nuclei. The global rms difference $\delta(\Delta B_{\infty}^{F})$ of infinite basis corrections to binding
energies between these two functionals  is $\delta(\Delta B_{\infty}^{F})=89.1$ keV. These two 
features will clearly result in unacceptable error in the calculations of binding energies. Thus,
the second ABOA round of optimization of the DD-MEZ functional is carried out using infinite
basis corrections to binding energies obtained for the DD-MEZ(round$_1$) functional. It
gives  the DD-MEZ(round$_2$) version for which infinite basis corrections are recalculated. 
This round leads to substantially smaller global rms difference of infinite basis corrections 
to binding energies obtained with  DD-MEZ(round$_1$) and DD-MEZ(round$_2$) versions of the 
functional being $\delta(\Delta B_{\infty}^{F})=16.3$ keV (see Fig.\ \ref{iter-proc})(b).  To achieve 
a higher precision in the calculation of infinite basis corrections to binding energies this iterative
procedure has to be repeated until the required precision $\delta(\Delta B_{\infty}^{F})$ 
is  achieved. We terminate this iterative procedure when  $\delta(\Delta B_{\infty}^{F})$
becomes smaller than 8 keV. This is achieved after two (for PC-Z) or three (NL5(Z) and
DD-MEZ [see Fig.\ \ref{iter-proc})(c)]) rounds of ABOA.  So, for example,  DD-MEZ(round$_3$) 
version of the functional is final and it becomes DD-MEZ CEDF. If higher level of accuracy is 
required additional rounds of  ABOA have to be carried out.

    The importance of this iterative procedure is clearly visible when infinite basis corrections 
to binding energies in the fermionic basis obtained for the Z  class of the functionals and those of
the DD-MEX, NL5(E) and PC-PK1 CEDFs are compared. These corrections are taken into account  
in the fitting protocol of the Z class of CEDFs and they are displayed in Fig.  3  of the present 
paper.  In contrast, they are neglected in the fitting protocols of the DD-MEX, NL5(E) and PC-PK1 
CEDFs since these functionals were fitted in  the $N_F=20$ fermionic and $N_B=20$ bosonic bases 
in Refs.\ \cite{PC-PK1,AAT.19,TAAR.20}.  However, afterwards infinite basis corrections to binding energies 
in the fermionic basis were calculated in Ref.\ \cite{TOAPT.24} and presented in Figs. 10(a), (c) and (g)
of this reference.

     Fig.\ \ref{iter-difference} compares infinite basis corrections to  binding energies 
in the fermionic basis obtained with the DD-MEZ and DD-MEX, NL5(Z) and NL5(E) and PC-Z and 
PC-PK1 pairs of the functionals.  One can see that there are global differences 
between these corrections obtained with the members of above mentioned pairs which are especially 
pronounced in actinides and superheavy nuclei.  Note that maximum difference of infinite basis 
corrections to  binding energies obtained with the members of these pairs reaches 0.381 MeV, 
0.265 MeV and 0.399 MeV for DDME, NLME and PC functionals, respectively.

\end{document}